 \documentclass[final,3p,times,twocolumn,authoryear]{elsarticle}

\usepackage{amssymb}
\usepackage{amsmath}
\usepackage{url}
\usepackage{algorithm}
\usepackage{algorithmicx}
\usepackage{algpseudocode}
\usepackage{diagbox}
\usepackage{float}

\journal{Journal of Systems and Software}

\begin{document}

\begin{frontmatter}



\title{CM-CASL: Comparison-based Performance Modeling of Software Systems via Collaborative Active and Semisupervised Learning}


\author[inst1]{Rong Cao\corref{cor1}}\ead{caorong@stu.xidian.edu.cn}
\author[inst1]{Liang Bao}\ead{baoliang@mail.xidian.edu.cn}
\cortext[cor1]{Corresponding Author}

\affiliation[inst1]{organization={School of Computer Science and Technology, Xidian University},
            addressline={No.2 South Taibai Road}, 
            city={Xi'an},
            postcode={710071}, 
            state={ShaanXi},
            country={China}}

\author[inst2]{Chase Wu}\ead{chase.wu@njit.edu}
\author[inst1]{Panpan Zhangsun}\ead{zhangsunpanpan@stu.xidian.edu.cn}
\author[inst1]{Yufei Li}\ead{yufeili@stu.xidian.edu.cn}
\author[inst1]{Zhe Zhang}\ead{420217317@qq.com}

\affiliation[inst2]{organization={Department of Data Science, New Jersey Institute of Technology},
            addressline={University Heights}, 
            city={Newark},
            postcode={07102}, 
            state={NJ},
            country={USA}}

\begin{abstract}
Configuration tuning for large software systems is generally challenging due to the complex configuration space and expensive performance evaluation. Most existing approaches follow a two-phase process, first learning a regression-based performance prediction model on available samples and then searching for the configurations with satisfactory performance using the learned model. Such regression-based models often suffer from the scarcity of samples due to the enormous time and resources required to run a large software system with a specific configuration. Moreover, previous studies have shown that even a highly accurate regression-based model may fail to discern the relative merit between two configurations, whereas performance comparison is actually one fundamental strategy for configuration tuning. To address these issues, this paper proposes CM-CASL, a Comparison-based performance Modeling approach for software systems via Collaborative Active and Semisupervised Learning. CM-CASL learns a classification model that compares the performance of two given configurations, and enhances the samples through a collaborative labeling process by both human experts and classifiers using an integration of active and semisupervised learning. Experimental results demonstrate that CM-CASL outperforms two state-of-the-art performance modeling approaches in terms of both classification accuracy and rank accuracy, and thus provides a better performance model for the subsequent work of configuration tuning.
\end{abstract}




\end{frontmatter}


\section{Introduction}
Software systems are becoming increasingly configurable for flexibility and adaptability \citep{Xu2015Hey, van2017automatic}. Generally, configuration parameters have a significant impact on the functional and non-functional properties of software systems. Oftentimes, performance is one of the most important non-functional properties as it directly affects user experience \citep{guo2018data}. Configuration tuning is to determine suitable configurations to optimize system performance, and has attracted a great deal of attention from both academia and industry. For a given software system, a straightforward approach is to measure the performance of all valid configurations and then identify the one that yields the best performance. However, this approach requires exhaustive search, which is practically infeasible because of the exponentially growing configuration space and the daunting cost of collecting performance measurements \citep{bao2019actgan,sarkar2015cost}. Note that performance evaluation typically requires real experiments in production systems. Such an evaluation is not only time-consuming (e.g., executing a large-scale, complex benchmark takes minutes to hours or even days) but also prohibitively expensive (e.g., running a real system in clouds for a short period of time may be charged for hundreds of dollars). Therefore, in practice, only a limited set of configurations can be measured, resulting in a scarcity of samples.

To solve this problem, learning-based configuration tuning has been studied in depth, which follows a two-phase process. In the first phase, a performance model is constructed using a small number of samples (i.e., different configurations and their corresponding performance measurements). In the second phase, promising configurations are identified by applying some search algorithms to the configuration space and then comparing their performance using the trained model. In this paper, we focus on the first phase, i.e., performance modeling, which is critical to configuration tuning as the accuracy of a performance model largely determines the tuning efficacy.

In many previous studies, performance modeling is formulated as a regression problem, and different regression-based models (e.g., neural network \citep{ha2019deepperf, mahgoub2017rafiki, zheng2014self}, Classification And Regression Trees (CART) \citep{guo2013variability, guo2018data, nair2018faster, sarkar2015cost, valov2015empirical, nair2017using}, Random Forest (RF) \citep{valov2015empirical, bei2015rfhoc}, etc.) are trained with a limited set of labeled samples from the configuration space. Such regression-based models can be used to predict the performance of any unseen configurations. However, due to the scarcity of samples, these models often suffer from low accuracy \citep{nair2017using}. Moreover, even a highly accurate regression-based model may not be able to discern the comparative relationship between the performance of two configurations \citep{chen2019regression}, as validated by our experimental results shown in Figure~\ref{compare} in Section~\ref{model}. However, during the search for the optimal configuration, what the search algorithms really need is to determine the comparative relationship between the performance of different configurations, rather than the specific performance values of configurations \citep{bao2018learning,nair2018finding,bei2015rfhoc,chen2015machine,tang2017system,wang2016novel,trotter2019forecasting,hua2018hadoop,bei2017mest,yu2018datasize}. Based on this observation, we propose to develop a comparison-based performance model \citep{bao2018autoconfig, zhu2019classytune} and formulate performance modeling as a classification problem. 

However, the comparison-based performance model still faces the challenge of sample scarcity. It is generally impractical to collect as many training samples as required for the model by actually running the software system because of the high monetary and temporal cost \citep{bao2019actgan}. Towards this end, we propose an active learning (AL) approach, where we can generate new training samples based on manual tuning experiences from human experts without actual running cost~\citep{zhu2019classytune}. During manual tuning, one would focus on the effect of a parameter change to the configuration on the performance (i.e., increase or decrease). Thus, the comparison-based model is more in line with the human way of thinking, and the tuning experiences of experts can be leveraged to label some new samples through comparison. AL provides a powerful tool to select high-quality samples for human experts to label, leading to an effective classifier \citep{tong2001active}. However, the labor-intensive manual labeling process in AL limits the number of training samples that can be added. To train a good classification model, the quality and the quantity of the training samples are equally important \citep{rajan2008active}. Semisupervised learning (SSL) improves the performance by utilizing unlabeled samples to increase the number of training samples, without human intervention \citep{zhu2003semi}. The key intuition is that some useful samples can be selected from the set of available unlabeled samples for labeling via collaborative active and semisupervised learning to enhance the quality and quantity of samples, further improve the comparison-based model.

In this paper, we propose a Comparison-based performance Modeling approach for configuration tuning of software systems via Collaborative Active and Semisupervised Learning (CM-CASL), where an integration of AL and SSL is adopted to perform collaborative labeling by both human experts and the classifier itself. AL achieves good prediction results by improving the quality of the training samples. It obtains high-quality samples by selecting effective samples iteratively using a query strategy for human labeling, which balances between informativeness and representativeness, thereby facilitating a deep fusion of the expert knowledge.  

Different from AL, SSL pays more attention to unlabeled samples in an unsupervised manner. It enhances the classifier by increasing the number of training samples, and improves the generalization of the classifier. However, the performance of SSL would depend on the informativeness and reliability of pseudolabeling. To address this issue, CM-CASL employs an AL-based verification scheme to improve the accuracy of pseudolabeling. Furthermore, the unlabeled samples at median distances are considered more informative and therefore are assigned with pseudolabels. As a result, the samples with more information and credible pseudolabels are added to increase the quantity of training samples, hence further improving the prediction performance.

In summary, our work makes the following contributions:
\begin{itemize}
\item We propose CM-CASL, which integrates AL and SSL to perform collaborative labeling by both human experts and classifiers, and provides a promising solution to comparison-based performance modeling.
\item We fuse the tuning experiences of human experts into a comparison-based performance model through AL without actual running cost, and thus improve the classification accuracy.
\item We evaluate the performance of CM-CASL through extensive experiments using 11 workloads in seven representative software systems, which show that CM-CASL outperforms two state-of-the-art baseline algorithms by 17.59\%-28.97\% for classification accuracy and 30.11\%-35.38\% for rank accuracy on average, yielding an average tuned performance improvement of 18.76\% and 16.05\%, respectively.
\end{itemize}

\section{Motivation}\label{motivation}
\subsection{A Comparison-based Performance Model}\label{model}
Performance comparison is one fundamental strategy for configuration tuning. Hence we propose to develop a model for performance comparison in support of configuration tuning. Specifically, given a software system with two different configurations $\emph{c}_{i}$ and $\emph{c}_{j}$ in the same running environment, we wish to build a comparison-based model, denoted as $\emph{f}: \mathbb{C}\times\mathbb{C}\rightarrow\{1,0\}$, which compares the performance values of the given software system with $\emph{c}_{i}$ and $\emph{c}_{j}$:
\begin{equation}
f(c_i,c_j)= \begin{cases}
1\quad &\text{if}\ f(c_i)-f(c_j)>0, \\
0\quad &\text{otherwise}.
\end{cases} 
\end{equation}
This comparison-based model takes a pair of configurations $(\emph{c}_{i},\emph{c}_{j})$ as input and outputs 1 if the former configuration has a performance better than the latter, or 0, otherwise.

The evaluation criterion of a comparison-based model is \emph{Classification Accuracy (CA)} \citep{novakovic2017evaluation}, which is defined as the percentage of correctly classified samples among all samples in the testing dataset:
\begin{equation}
CA = \frac{N_{\text{correctly\ classified\ samples}}}{N_{\text{all\ samples}}}\cdot 100\%.
\end{equation}
For a regression-based model, we use \emph{Mean Relative Error (MRE)} as a metric to evaluate prediction accuracy \citep{guo2018data, ha2019deepperf}, which is computed as:
\begin{equation}
    MRE=\frac{1}{N}\sum_{i=1}^{N}\frac{|a_i-p_i|}{a_i}\cdot 100\%,
\end{equation}
where $N$ is the size of the testing dataset, and $a_i$ and $p_i$ represent the actual performance and the predicted performance, respectively.

\begin{figure}[h]
\centering
\includegraphics[width=0.9\linewidth]{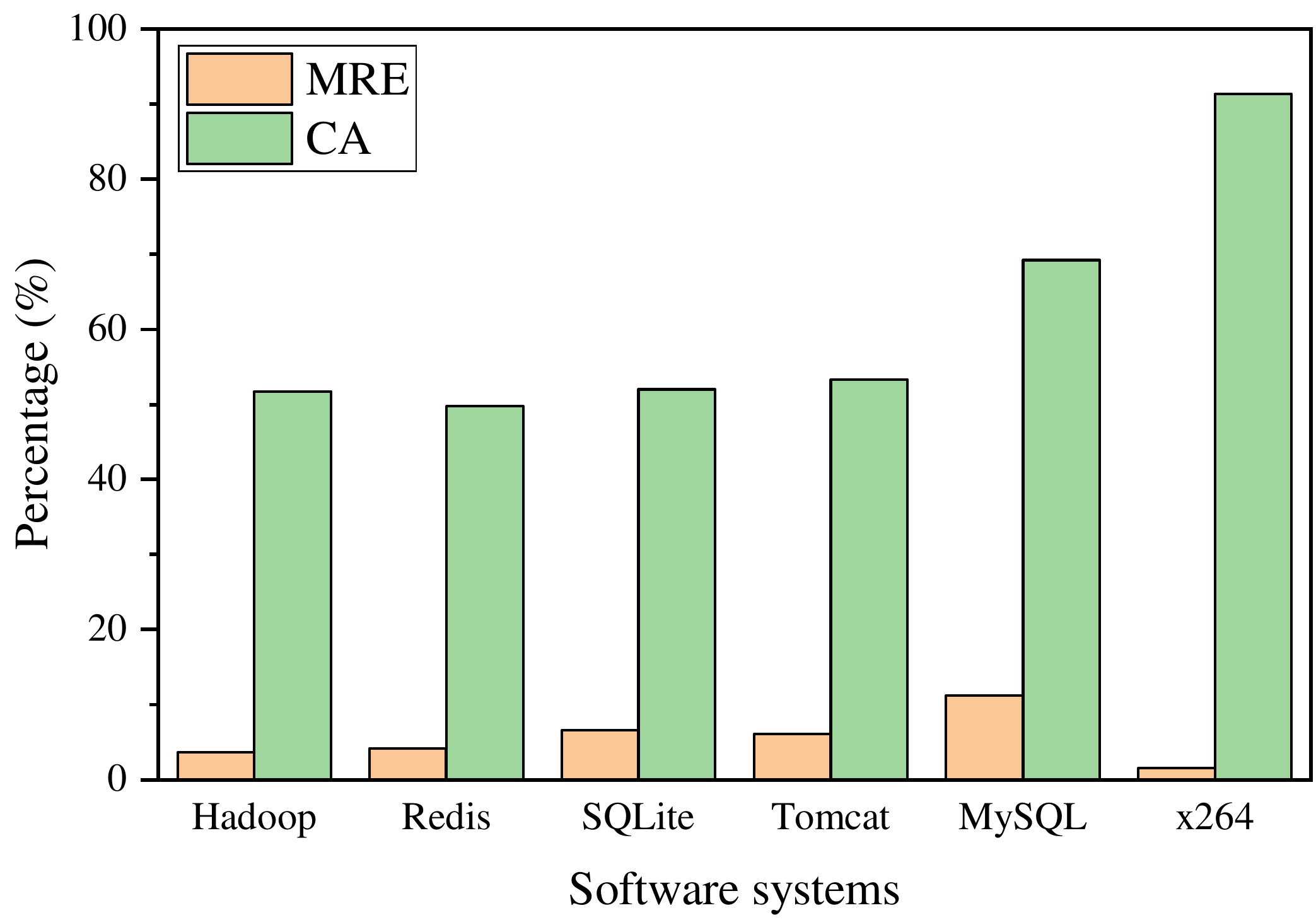}
\caption{The comparison between MREs and CAs of different software systems obtained by DeepPerf.}
\label{compare}
\end{figure}

Several of such regression models have achieved high prediction accuracy, which, however, may not be sufficient to correctly predict the relative merit between two configurations. For instance, consider two configurations $\emph{c}_{1}$ and $\emph{c}_{2}$, which result in a performance of 150 and 160, respectively, and a regression model predicts the performance of $\emph{c}_{1}$ and $\emph{c}_{2}$ to be 155 and 154, respectively. The MRE of this model is only 3.54\%, but it\ incorrectly predicts $\emph{c}_{1}$ to be a better configuration than $\emph{c}_{2}$. To demonstrate this point, we compare the MREs and CAs of different software systems obtained by DeepPerf (a state-of-the-art regression-based model) \citep{ha2019deepperf}, and the results are shown in Figure~\ref{compare}. The comparison results are determined by the predicted performance of any two configurations. Ideally, the lower the MRE, the higher the CA, as in x264. However, low MREs do not always yield the desirable CAs for the complex systems in the experiment.

In addition, employing a comparison-based model has the following advantages. First, the classification model for performance comparison can increase the number of available training samples, as it takes pairs of original samples as input and such comparison-based samples can be constructed by combining every pair of the original samples. Second, we can generate even more training samples based on manual tuning experiences of human experts, without running cost. The manual tuning process usually involves a series of trials and errors, thus resulting in some comparison-based empirical rules. We can generate more training samples for the classification model by inquiring the experts which of the two configurations achieves better performance, while this is impossible for the regression-based performance modeling.

\subsection{Sample Enhancement with AL and SSL}\label{sample}
Limited samples caused by expensive measurements are another barrier to achieving accurate performance comparison. Generally, labeled training data are very limited, but unlabeled data are abundant or essentially unlimited. Therefore, a straightforward idea is to use the unlabeled data to enhance the training data. AL and SSL are two effective approaches to address such a problem. AL needs human involvement and aims at selecting the most useful samples to label for training, but the labor-intensive manual labeling process limits the number of training samples that can be added. Conversely, SSL assigns pseudolabels to unlabeled samples without human involvement, but it may cause wrong pseudolabeling and further degrade classification accuracy. To alleviate the above problems, we integrate the AL and SSL together to train our comparison-based model using a collaborative labeling process by both human experts and classifiers.

\section{Our Approach}\label{approach}
\begin{figure}[h]
  \centering
  \includegraphics[width=\linewidth]{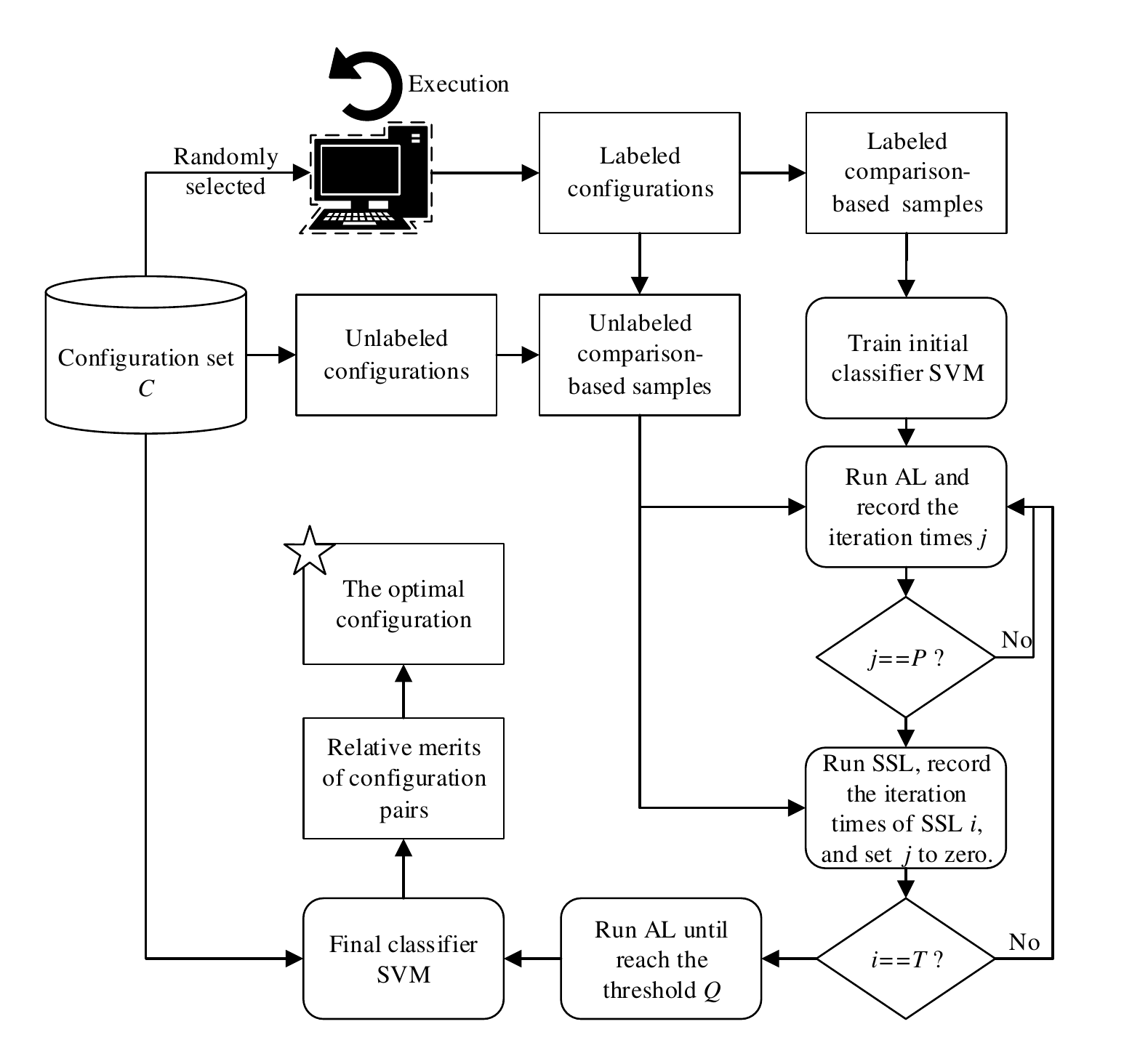}
  \caption{The modeling process of CM-CASL. After each successive $P$ iterations of AL, an iteration of SSL is performed. After $T$ iterations of SSL, the modeling process terminates once the number of iterations of AL reaches the threshold $Q$. Otherwise, AL process is continued. The final output is a classifier that can discern the relative merit between any two configurations.}
  \label{process}
\end{figure}

In this paper, we propose CM-CASL, a comparison-based performance modeling approach for configuration tuning of software systems via collaborative active and semisupervised learning, as illustrated in Figure~\ref{process} and summarized below.

\textbf{Data collection}. We consider a configuration set with $n$ samples $C = \{c_1, c_2, \cdots, c_n \}$, where $c_i$ denotes the \emph{i}-th configuration with $d$ parameters of the software system. We first randomly select $l$ samples from $C$, and obtain their performance. The remaining $u=n-l$ samples are configurations without measurements. Then, we generate a comparison-based sample set $S$ by combining every pair of the samples in $C$, and assign a label to each sample, depending on the performance of two configurations in this pair. The ${l\choose 2}$ labeled comparison-based samples form $S_L$, while the rest of the comparison-based samples are considered as the unlabeled sample set $S_U$.

\textbf{Comparison-based modeling}. Given the limited availability of initial labeled samples, CM-CASL starts with an initial classifier SVM, and employs AL and SSL collaboratively to get more promising training samples, and uses them to enhance the performance of the final classifier. The final SVM classifier is able to compare two configurations with higher accuracy, further provides support for configuration tuning of software systems. 

\subsection{Query Strategy for AL}\label{query}
AL denotes the process of autonomously selecting promising data points to learn from \citep{kremer2014active,settles2009active}, and it is suitable for solving classification problems with limited labeled samples. An initial classifier is first trained with a small number of samples, and then some promising samples are selected by a carefully designed query strategy for human experts to label. These newly obtained training samples are used to update the classifier, resulting in a better classifier. Moreover, we are confronted with large amounts of unlabeled data, so estimating the effect of single samples is costly. A speedup can be gained by labeling samples in batches. Another reason for batch labeling is the demand for the SSL framework to maintain differences between models before and after each AL iteration.

The query strategy is the key to batch-mode AL. The most commonly used query strategy is uncertainty sampling. In this strategy, 
AL selects those unlabeled samples that are difficult to predict by the current classifier to query \citep{lewis1994sequential}. In the case of an SVM, it is intuitive to query the labels of the samples that are closest to the decision boundary in each iteration \citep{tong2001active,tong2001support}. Uncertainty sampling is a feasible strategy to utilize informativeness, aiming to reduce the uncertainty of current classifier.

However, for batch-mode AL, this strategy does not take into account that the multiple samples queried in a single iteration may have outliers or be similar. This may cause the AL process to focus too much on certain regions while ignoring some more representative ones. To avoid this problem, we need to consider a combination of informativeness and representativeness when selecting the samples to query \citep{dasgupta2011two,settles2009active}.

To avoid oversampling unrepresentative outliers or similar samples, we combine uncertainty sampling and clustering \citep{xu2003representative}. Considering the weak performance of our classifier due to the small initial samples, we design a cluster-based uncertainty sampling scheme for the query strategy of batch-mode AL. First, an SVM classifier is trained using the labeled samples $S_L$ (line 1). To ensure the representativeness of the samples to query, we perform $k$-means clustering on all unlabeled samples to obtain the medoids of the $k$ groups (line 2-3). Finally, to ensure the informativeness of the samples, we query the $q=k/n$ medoids that are closest to the decision boundary of the current SVM classifier (line 5-6). Since only the medoids closest to the decision boundary are considered, these samples all have a high level of uncertainty, and the value of $n$ can adjust the degree of uncertainty. Clustering, on the other hand, increases the informativeness by excluding samples that are too similar. This way, our query strategy achieves a good balance between informativeness and representativeness, as detailed in Algorithm ~\ref{alg:QS}.
\begin{algorithm}[h]
      \caption{Cluster-based Uncertainty Sampling}
      \label{alg:QS}
      \begin{algorithmic}[1]
        \Require
          $S_L$: labeled samples;
          $S_U$: unlabeled samples;
          $n$: the parameter that adjusts the degree of uncertainty;
          $q$: the batch size in AL.
        \Ensure
          $S_Q$: samples selected to be queried.
        \State Train SVM using $S_L$;
        \label{code:QS:model}
        \State $k$ $\leftarrow$ $q \times n$;
        \label{code:QS:k}
        \State $k$-clustered $S_U$ $\leftarrow$ $k$-$means(S_U, k)$;
        \label{code:QS:kmeans}
        \State $k$ medoids in $S_U$ $\leftarrow$ $k$-clustered $S_U$;
        \label{code:QS:medoids}
        \State $S_K$ $\leftarrow$ $k$ medoids in $S_U$;
        \label{code:QS:Smedoids}
        \State $S_Q$ $\leftarrow$ $\underset{x_i}{min} \{d(x_i)|x_i \in S_K\}$;
        \label{code:QS:SQ} \\
        \Return $S_Q$;
      \end{algorithmic}
\end{algorithm}

\subsection{Pseudolabel Assignment and Verification in SSL}\label{pseudolabel}
Some SSL algorithms increase the number of training samples of the classifier by assigning pseudolabels to unlabeled samples \citep{yarowsky1995unsupervised,blum1998combining,zhou2005tri}. It first trains the classifier with labeled samples, and uses the trained classifier to predict pseudolabels for unlabeled samples, then retrain the classifier with the pseudolabeled and labeled samples together. One key advantage of pseudolabeling is the ability to generate more training samples for classifier to enhance the performance and robustness. 

However, the strategies for pseudolabel assignment have to be carefully designed. The most intuitive approach is to assign pseudolabels to samples that the classifier can predict with high confidence. But high classification confidence means low information, often leading to limited effects on model performance. Therefore, we propose to select unlabeled samples at median distances to the current decision boundary to assign pseudolabels. Without prior knowledge about the distribution of the classes, it is reasonable to consider that the unlabeled samples at median distances are at the center of the classes, which are more informative to the classifier.

In addition, incorrect pseudolabeling is also a problem that can easily occur. Subsequently, more and more incorrect pseudolabels may be introduced during the training, worsening the accuracy of classifier \citep{bruzzone2010recent}. To solve this problem, we employ an AL-based pseudolabel verification process to ensure labeling reliability.

The pseudolabels of unlabeled samples are verified in successive iterations of AL. When the predicted label of an unlabeled sample $x_i \in S_U$ does not change during successive $P$ iterations of AL, the sample is considered as having been correctly classified. The predicted label of $x_i \in S_U$ in each iteration is denoted as $L_t(x_i)$, where $t$ represents the $t$-th iteration. The number of label changes $N_{change}$ of $x_i \in S_U$ during the successive $P$ iterations can be calculated as:
\begin{equation}
    N_{change}(x_i)=\sum_{t=j}^{j+P-1}CH_t(x_i),
\end{equation}
where
\begin{equation}
CH_t(x_i)= \begin{cases}
1\quad &if\ L_t(x_i) \neq L_{t-1}(x_i), \\
0\quad &if\ L_t(x_i) = L_{t-1}(x_i).
\end{cases} 
\end{equation}

During the iterations of AL, the decision boundary would be continuously adjusted, thus the predicted labels of some unlabeled samples may change in each iteration. If $N_{change}$ of an unlabeled sample is equal to zero, it indicates that the predicted label of this sample has not changed during the adjustment of the decision boundary. In this case, we consider that this sample has been correctly classified with high probability.

We integrate into the SSL framework the pseudolabel assignment for unlabeled samples at median distances and the pseudolabel verification mechanism via $N_{change}$. For the comparison-based classification model, during the execution of AL, whenever AL iterates $P$ times, the unlabeled samples can be classified into the positive class $S_P$ and negative class $S_N$ by the current classifier.

Before assigning pseudolabels, we need to conduct a pseudolabel verification process. The values of $N_{change}$ for all remaining unlabeled samples in both positive and negative classes are computed. At first, we identify the unlabeled samples whose $N_{change}$ is equal to zero. By using $N_{change}$ to ensure the reliability of the pseudolabels, sample sets $R_P$ and $R_N$ would have higher classification confidence on the pseudolabels of these unlabeled samples.

Then, the unlabeled samples at median distances to the decision boundary are selected from the positive class and the negative class, respectively. For each class, the samples in $X_P$ and $X_N$ at median distances to the decision boundary would not only have credible pseudolabels, but also be more informative. By this pseudolabel assignment and verification mechanism, we increase the quantity of training samples for classifier efficiently. 

\subsection{An Elaboration of CM-CASL}\label{detail}
The abovementioned SSL algorithm relies on AL to verify the reliability of predicted pseudolabels, while the added training samples assigned with pseudolabels are informative to AL. Thus, AL collaborates with SSL to improve the classifier. The pseudocode of CM-CASL is provided in Algorithm~\ref{alg:CM-CASL} and explained in detail.

Considering the limited query opportunities of AL, we set the termination condition for CM-CASL to be the number of the expanded labeled samples by AL exceeding a threshold $Q$. Consequently, we set the iteration number of AL and SSL according to the batch size in AL, the successive iteration times of AL, and the threshold value (line 1-2). At the beginning of CM-CASL, we use the labeled samples $S_L$ to train the classifier SVM (line 3). The successive $P$ iterations of AL are performed subsequently. In each iteration of AL, $q$ informative and representative samples $S_Q$ are chosen from $S_U$ based on the trained SVM and the cluster-based uncertainty sampling (line 6). We update $S_U$ and $S_L$ by removing $S_Q$ from $S_U$ and adding $S_Q$ with human labeling to $S_L$ (line 7-8). After that, we use the newly labeled dataset $S_L$ to retrain the classifier (line 9), and obtain the classification results of unlabeled samples $S_U$ using the current classifier (line 10).

With the classification results of $S_U$ in the successive iterations of AL, $N_{change}$ of the remaining unlabeled samples can be calculated (line 12). The samples with reliable predicted results are selected via $N_{change}$ that is equal to zero (line 13). Within these credible samples, we suppose that the samples at median distances to the decision boundary are more informative about the data distribution (line 14). The selected $X_P$ and $X_N$ are assigned with pseudolabels (line 15), and meanwhile, the datasets $S_U$ and $S_L$ are updated (line 16). The added samples with pseudolabels are also applied to update the SVM classifier (line 17). The remaining unlabeled samples in $S_U$ that cannot be assigned with pseudolabels are treated as the candidates for AL to launch the next iteration of SSL. The above process is repeated in an iterative manner.
\begin{algorithm}
      \caption{CM-CASL}
      \label{alg:CM-CASL}
      \begin{algorithmic}[1]
        \Require
          $S_L$: labeled samples;
          $S_U$: unlabeled samples;
          $P$: the successive iteration times of AL;
          $q$: the batch size in AL;
          $t$: the batch size in SSL;
          $Q$: the threshold for the number of the expanded labeled samples by AL.
        \Ensure
          $f$: the SVM classifier.
        \State Calculate the iteration times of AL: $M=Q/q$;
        \State Calculate the iteration times of SSL: $T=M/P$;
        \State Train SVM using $S_L$;
        \For{$i=1$ to $T$}
        \For{$j=1$ to $P$}
        \State Select $q$ promising samples $S_Q$ from $S_U$ based on SVM and Algorithm~\ref{alg:QS};
        \State Label $S_Q$ by human experts;
        \State Update $S_L=S_L \cup S_Q$ and $S_U=S_U \setminus S_Q$;
        \State Update SVM using new $S_L$;
        \State Classify $S_U$ using $SVM_j$ with the results being $Label_j$;
        \EndFor
        \State Calculate $N_{change}$ for $S_U$;
        \State Select $R_P$ and $R_N$ with $N_{change}=0$ from $S_P$ and $S_N$, respectively;
        \State Select $t/2\ X_P$ and $t/2\ X_N$ with median distances to the decision boundary from $R_P$ and $R_N$;
        \State Assign pseudolabels to $X_P$ and $X_N$;
        \State Update $S_L=S_L \cup X_P \cup X_N$ and $S_U=S_U \setminus (X_P \cup X_N)$;
        \State Update SVM using new $S_L$;
        \EndFor
        \While {$q\times P\times T < Q$}
        \State Select $q$ promising samples $S_Q$ from $S_U$ based on SVM and Algorithm~\ref{alg:QS};
        \State Label $S_Q$ by human experts;
        \State Update $S_L=S_L \cup S_Q$ and $S_U=S_U \setminus S_Q$;
        \State Update SVM using new $S_L$;
        \EndWhile\\
        \Return SVM;
      \end{algorithmic}
    \end{algorithm}

After $T$ iterations of SSL, the algorithm terminates once the number of expanded labeled samples reaches a predefined threshold $Q$. Otherwise, it continues the AL process until the termination condition is satisfied (line 19-24). 

\section{Performance Evaluation}\label{experimental}
We implement CM-CASL and other algorithms, and conduct extensive experiments in diverse software systems. The source code and the data can be found in the online anonymous repository: \url{https://github.com/xdbdilab/CM-CASL}. In this section, we first describe our experiment setup, and compare the performance of CM-CASL with that of two state-of-the-art baseline algorithms. Then, we evaluate the impact of AL and SSL on the performance of CM-CASL, respectively. After that, the sensitivity of CM-CASL to expert accuracy is evaluated. Finally, we evaluate whether CM-CASL can improve the efficacy of configuration tuning.

\subsection{Experimental Methodology}\label{methodology}
\textbf{Subject Software Systems, Benchmarks, Parameters, and Performance Metrics}. We choose seven widely used software systems to evaluate CM-CASL, which have different characteristics and originate from different application domains, e.g., data analytics, database management systems, web servers, video encoders, etc. The information on the selected software systems is shown in Table \ref{tab:subject}. Specifically, Hadoop is a framework that allows for distributed processing of large datasets across clusters of computers, Spark is a cluster computing engine for big-data analytics, MySQL is an open-source relational database management system (RDBMS), SQLite is an open-source embedded RDBMS, Redis is an open-source in-memory data structure store, Tomcat is an open-source implementation of Java Web technologies, and x264 is a video encoder that encodes raw videos into the H.264 compressed format.
\begin{table*}[htbp]
  \caption{Subject systems, benchmarks, parameters, and performance metrics.}
  \centering
  \label{tab:subject}
  \resizebox{0.9\textwidth}{!}{
  \begin{tabular}{lllcl}
    \hline
    Subject systems &Category  &Benchmark & \# of selected parameters &Performance\\
    \hline
    Hadoop    & Data analytics  & HiBench     & 9 & Throughput (MB/s)\\
    Spark     & Data analytics  & HiBench     & 13& Throughput (MB/s)\\
    MySQL     & RDBMS      & sysbench    & 10& Latency (ms)\\
    SQLite    & Embedded DB     & Customized  & 22 & Transactions per second\\
    Redis     & In-memory DB    & Redis-Bench & 9 & Requests per second\\
    Tomcat    & Web server    & JMeter      & 14& Requests per second\\
    x264      & Video Encoder  & Customized  &  9& kilobit per second (kb/s)\\
    \hline
  \end{tabular}}
  \end{table*}

For each software system, the performance is measured using a standard benchmark, either delivered by its vendor or used widely in its application domain. We use HiBench \citep{huang2010hibench} for Hadoop and Spark,  sysbench\footnote{https://github.com/akopytov/sysbench} for MySQL, Redis-Bench\footnote{https://redis.io/topics/benchmarks} for Redis, JMeter\footnote{https://jmeter.apache.org/} for Tomcat, and customized workloads for SQLite and x264 respectively.

In addition, we use domain expertise to identify a subset of parameters that are considered critical to the performance, as in \citep{zhu2017bestconfig,bao2019actgan,bei2015rfhoc,ha2019deepperf}. Note that reducing the number of considered parameters can reduce the configuration space exponentially, and numerous existing tuning approaches \citep{bao2019actgan,zhu2017bestconfig,sarkar2015cost,bei2015rfhoc} adopt this manual feature selection strategy.

\textbf{Running Environment}. To avoid interference in collecting samples from different software systems, we conduct experiments sequentially on different cloud servers. The running environments for different systems are as follows.

Hadoop and Spark experiments are conducted on a cluster of three cloud servers. Each server is equipped with two 4-core Intel(R) Xeon(R) CPU E5-2682 v4 @2.50GHz processors, 8GB RAM, and a 100GB disk, and is installed CentOS 7.6 and Java 1.8.0.

MySQL, SQLite, Redis, Tomcat, and x264 are run on a cloud server equipped with two 2-core Intel(R) Xeon(R) CPU E5-2682 v4 @2.50GHz processors, 4GB RAM, and a 50GB disk, and is installed CentOS 7.6 and Java 1.8.0.

\textbf{Baseline Algorithms}. To evaluate the performance of CM-CASL, we compare it with two state-of-the-art algorithms, namely, DeepPerf \citep{ha2019deepperf} and rank-based CART \citep{nair2017using}. DeepPerf treats the performance prediction problem as a non-linear regression problem and uses a deep feedforward neural network (FNN) combined with L1 regularization to model software systems. The key idea of rank-based CART (denoted as R-CART) is to use ranking as an approach for building regression models. A CART model, which may not be accurate but preserves the rank information of configurations, is trained to find good configurations. The hyperparameters of these baseline algorithms are optimized using grid search strategies.

\textbf{Evaluation Metrics}. We consider two performance metrics in our experiments for performance evaluation, namely, classification accuracy (CA) and rank accuracy (RA). 

The most direct evaluation criterion of the classification model for performance comparison is CA, which is defined in Section~\ref{model} as the percentage of correctly classified samples to all samples in the test dataset.

Moreover, we introduce RA to evaluate the rank performance of different models. Once a model is trained, the ranking of $N$ different configurations can be generated by simply using some sorting algorithm and the comparison-based model. After obtaining the predicted rank order, it is compared to the actual ranking order. The RA is thus calculated as the mean rank difference \citep{bao2018autoconfig,nair2017using}:
\begin{equation}
    RA=\frac{1}{N}\sum_{i=1}^{N}|rank(y_i)-rank(P(c_i))|,
\end{equation}
where $P(c_i)$ denotes the predicted performance of the $i$-th configuration ($i=1,\cdots,N$), and $y_i$ denotes the actual performance of the $i$-th configuration. To compare the prediction accuracy of CM-CASL with that of the two baselines, the CA and RA of CM-CASL on the test data are normalized to that of the DeepPerf approach.

\textbf{Experimental Settings}. In practice, the time for configuration tuning and performance modeling is often restricted \citep{bao2019actgan,zhu2017bestconfig}. We define the restricted modeling time as time constraint (TC), and set the TC to be a fixed value in this work. Each modeling approach must complete within the preset TC.

DeepPerf and R-CART collect all samples by performance measurements, while CM-CASL uses part of the time for performance measurements in the system and the rest for manual labeling. The samples collected by performance measurements can be randomly sampled or selected by human experts. 

The testing dataset is constructed by randomly selecting 50 configuration-performance pairs in each subject system to evaluate the RA of different approaches, and comparison-based samples are generated using the 50 configuration-performance pairs for the evaluation of CA.

\textbf{Manual Labeling in CM-CASL}. We can generate new training samples based on manual tuning experiences from human experts without actual running the system. In this way, we bring in expert knowledge through comparison-based samples. A manual labeling task is to give two configurations of a system and let a human expert decide which configuration may have better performance.

\textbf{Preliminary Experiment of CM-CASL}. To evaluate the performance of CM-CASL, we recruit 20 participants (6 female, 14 male; age M=32.1 and SD=4.57) with experience in tuning and optimizing complex software systems ranging from one year to more than ten years. The participants are divided into different groups according to their familiarity with different software systems, and the different groups are experimented with their familiar systems.

In the preliminary experiment, participants are asked to label a number of comparison-based samples randomly selected from one of their familiar systems. To stress test the manual labeling, we set the initial number of manual labeling to 300 and record the labeling time of each sample, while calculating the labeling accuracy. We first analyze the trend of labeling accuracy as the number of labeling rises. The results show that for most participants, the labeling accuracy decreases significantly when the number of labeling is greater than 200. We believe this is due to the fact that the continuous manual labeling can be extremely exhausting for the participants, so the subsequent analysis is conducted only for the data collected from the first 200 labeled samples for all participants.

We observe that the difference in labeling time between participants is small and does not change significantly as the number of labeling rises (labeling time M=29.7s and SD=7.89s). The reasons for the relatively small labeling time are twofold. First, the participants have rich tuning experience and domain knowledge, and our approach allows performance measurements for some initial samples in actual experiments, which further enhances the participants' knowledge of the subject system. Second, for each software system, a small number of important parameters are selected among a large number of parameters by domain knowledge. Participants often have a deep understanding of the impact of these parameters on performance, thus enabling fast labeling of the samples.

In addition, there is a strong correlation between the labeling accuracy and the working years (denoted as $T_w$) of the participants, which can be roughly divided into three levels: (1) when $T_w<5$, the labeling accuracies of participants (three in total) are usually between 0.7 and 0.8; (2) when $5\leq T_w<10$, the labeling accuracies of participants (15 in total) tend to be between 0.8 and 1; and (3) when $T_w\geq10$, the labeling accuracies of participants (two in total) are generally greater than 0.9. Moreover, participants are unable to discern the relative merit between two configurations in rare cases. The probability of this scenario occurring is less than 5\%. Therefore, in this scenario, we obtain the performance by actually running the software system, and label the comparison-based samples without adding much time cost.

\textbf{Experimental Procedure of CM-CASL}. Firstly, we select the largest number and most prevalent participants (i.e., those with working years between five and ten) to conduct experiments with the CM-CASL algorithm to demonstrate the efficacy of our proposal. Then, we conduct experiments on different variants of CM-CASL to explore the impact of AL and SSL on the performance of CM-CASL. The experiments here are set up in the same way as the above experiments. After that, we perform a sensitivity analysis of CM-CASL on labeling accuracy, which is achieved by selecting participants with different labeling accuracies for the same system among participants with different working years. Finally, we evaluate whether CM-CASL can improve the tuning efficacy by providing a better performance model. The performance modeling procedures of the above experiments are almost the same, and the differences mainly lie in the different parameter settings and algorithm variants, which have no influence on the participants' perspective. Therefore, we focus on the procedure of the first experiment here. Moreover, we ensure that the performance evaluation for each workload is performed by three participants, which means that each expert may be assigned to more than one group. The results are presented as the average of three experiments involving different participants.

Since the time for configuration tuning and performance modeling is usually limited, this work mainly considers the time costs of different approaches. The time cost for collecting a performance measurement varies across subject systems and workloads, ranging from one minute to ten minutes on average in our experiments. In the preliminary experiment, we observe that the average time cost for manual labeling is about 30 seconds, substantially lower than a performance measurement in the actual system, and the number of samples that most participants can label accurately and continuously is 200. Considering the different selection ratios and time costs, we set two TCs in our experiments, i.e., 2h and 3h. Furthermore, it is worth mentioning that in real production systems, especially in big data scenarios, the time cost of collecting a performance measurement varies from minutes to hours \citep{babu2009automated,bei2015rfhoc,mahgoub2017rafiki}. Since the time cost of manual labeling by human experts is relatively stable, a larger time cost for performance measurements would bring more potential for CM-CASL.

For 11 workloads in seven subject software systems, each participant is asked to complete six modeling tasks with the help from CM-CASL. The six modeling tasks for each workload follow a $2\times3$ design along two factors: modeling time constraints (2h and 3h) and selection ratios (1/2, 1/3, and 1/4). We conduct experiments with different selection ratios to demonstrate the robustness of CM-CASL. Therefore, the total experimental time for each workload is about 15 hours, and each participant performs experiments for one or two workloads depending on the grouping. All experiments are completed within one month.

Before a participant starts to work on any modeling tasks, we explain the comparison-based performance modeling task to the participant and walk through an example of the CM-CASL algorithm with step-by-step instructions. Participants are also given basic instructions on how to manual labeling the comparison-based samples selected by each AL iteration, and completing different modeling tasks with different settings. To ensure that all participants start different modeling tasks in each workload with the same amount of knowledge, the participants receive no feedback on whether their predictions for these unlabeled samples are correct in different modeling tasks with the same workload.

In CM-CASL, the threshold $Q$ is determined by the participants themselves according to TC and selection ratio, typically between 100 and 200. Moreover, we set the batch size in AL (i.e., $q$) to be 10 in the experiment. The iteration times of AL and SSL can be calculated accordingly. Besides, the successive iteration times of AL (i.e., $P$) and the batch size in SSL (i.e., $t$) are optimized by grid search through several experiments.

For each modeling task, participants randomly select a number of configurations and obtain the corresponding performance at first. The performance measurements are completed in a given amount of time. These labeled samples are used to train an initial SVM classifier. Subsequently, in each AL iteration, participants are asked to label some comparison-based samples selected by the query strategy for AL, i.e., the cluster-based uncertainty sampling. These manual labeled samples are used to update the classifier and recommend useful unlabeled samples in the subsequent AL iterations.

The main job of participants is to label the selected unlabeled samples during the AL iterations, while the SSL process requires no participant intervention. The output of CM-CASL is the final classifier, which can discern the comparative relationship between the performance of two configurations. The CA and RA of final classifier are compared with two baselines to verify the efficacy of CM-CASL.

\subsection{Experimental Results}\label{results}
\begin{figure}[t]
  \centering
  \includegraphics[width=\linewidth]{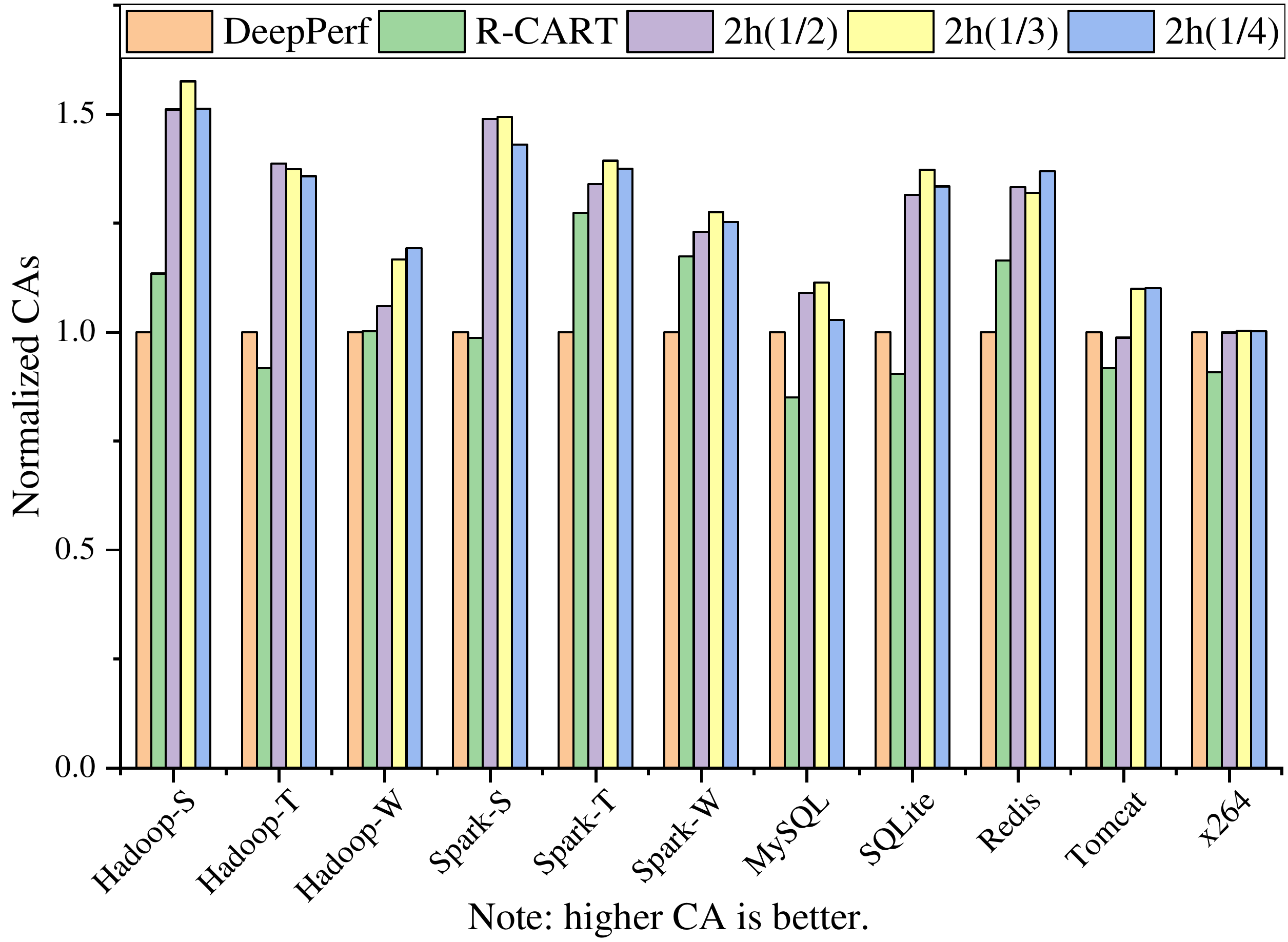}
  \caption{Normalized CAs of DeepPerf, R-CART, and different settings of CM-CASL within 2h (normalized to the corresponding CA of DeepPerf).}
  \label{2h-CA}
\end{figure}

\begin{figure}[h]
  \centering
  \includegraphics[width=\linewidth]{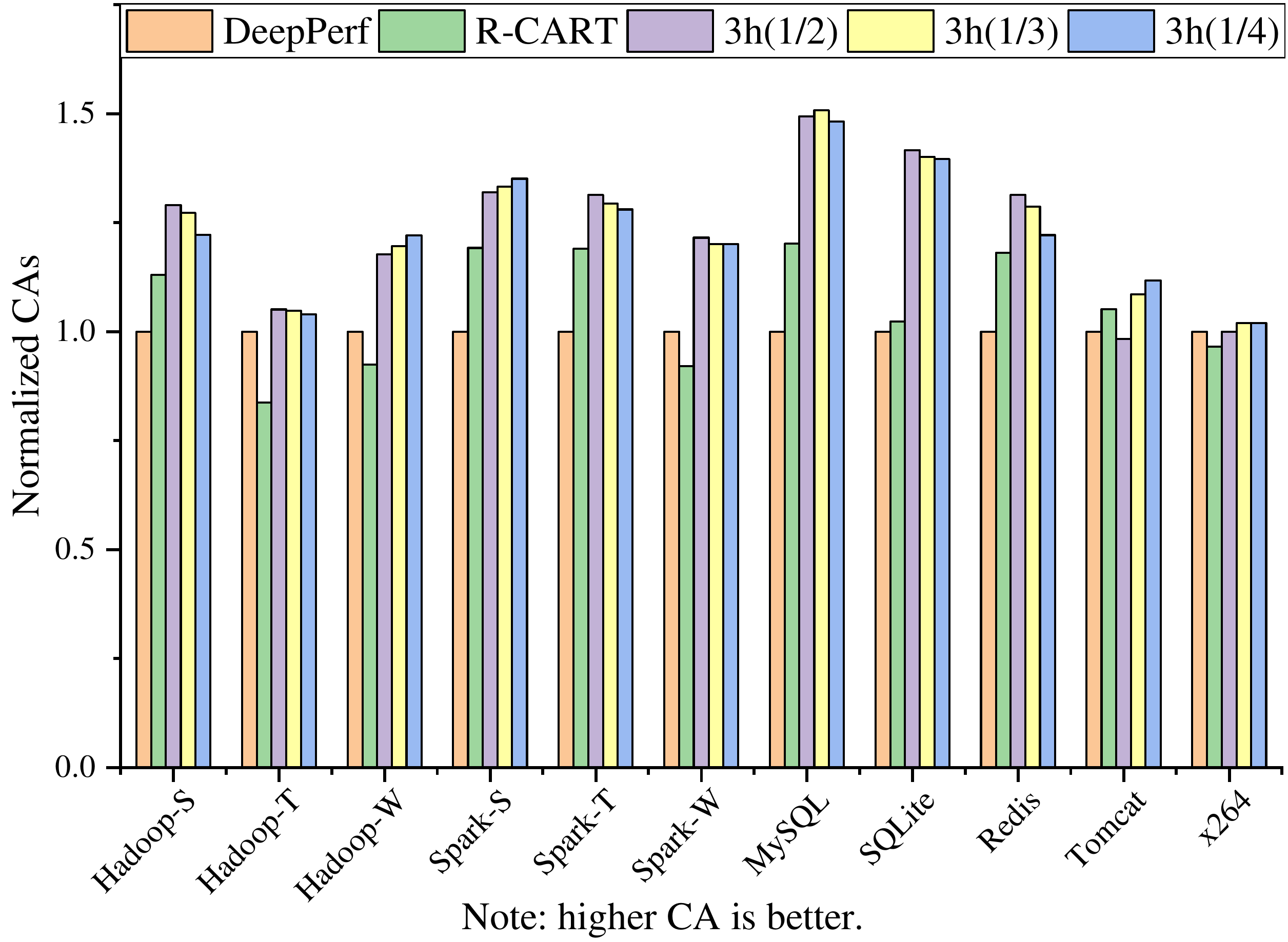}
  \caption{Normalized CAs of DeepPerf, R-CART, and different settings of CM-CASL within 3h (normalized to the corresponding CA of DeepPerf).}
  \label{3h-CA}
\end{figure}

\begin{figure}[h]
  \centering
  \includegraphics[width=\linewidth]{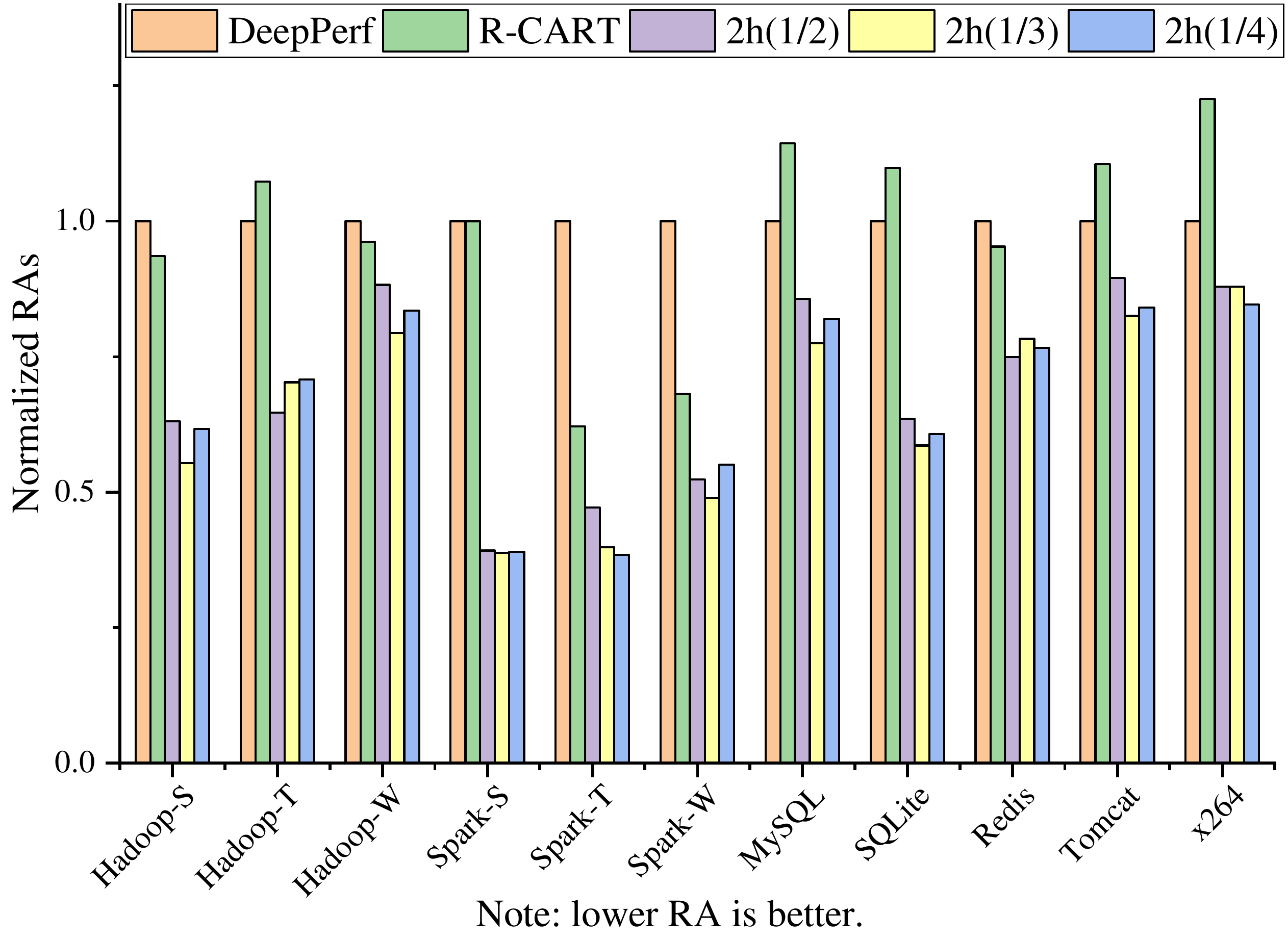}
  \caption{Normalized RAs of DeepPerf, R-CART, and different settings of CM-CASL within 2h (normalized to the corresponding RA of DeepPerf).}
  \label{2h-RA}
\end{figure}

\begin{figure}[h]
  \centering
  \includegraphics[width=\linewidth]{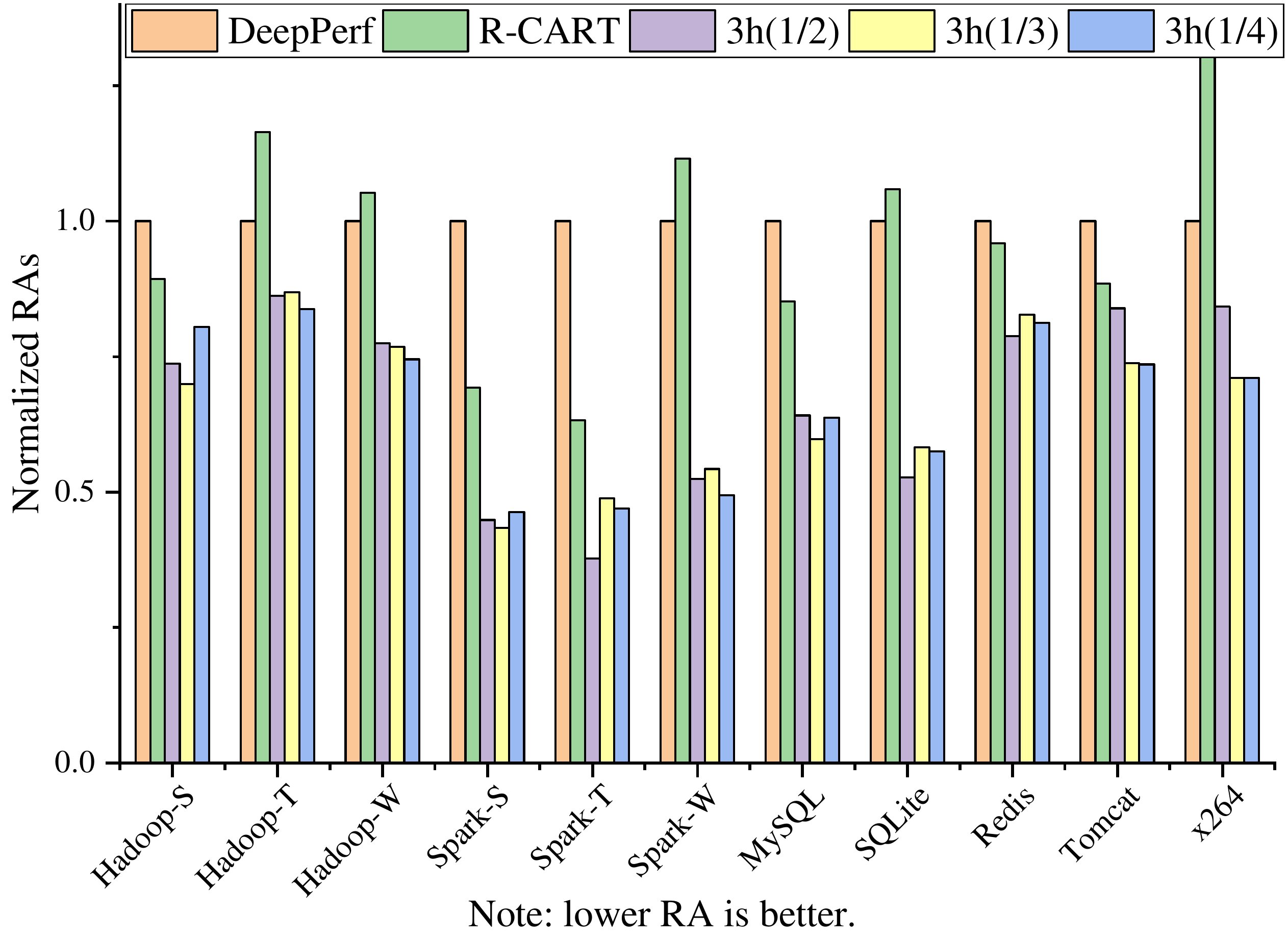}
  \caption{Normalized RAs of DeepPerf, R-CART, and different settings of CM-CASL within 3h (normalized to the corresponding RA of DeepPerf).}
  \label{3h-RA}
\end{figure}

The normalized CAs of DeepPerf, R-CART, and different settings of CM-CASL within different time budgets (2h and 3h) are shown in Figure~\ref{2h-CA} and Figure~\ref{3h-CA}, while the comparison of normalized RAs within different time budgets (2h and 3h) are shown in Figure~\ref{2h-RA} and Figure~\ref{3h-RA}, respectively. Note that $S$, $T$, and $W$ represent three different workloads, namely, Sort, Terasort, and Wordcount in the figures.

We observe that CM-CASL outperforms other approaches in different subject systems and workloads. On average, CM-CASL increases the CAs of DeepPerf by 26.91\% and 23.55\% within 2h and 3h, respectively. Moreover, CM-CASL reduces the RAs of DeepPerf by 33.04\% and 33.62\%  with the labeling time being 2h and 3h, respectively. Most notably, CM-CASL reduces the RAs of DeepPerf by 47.91\%-61.03\% under different Spark workloads. Compared with R-CART, CM-CASL improves the CAs by 25.28\% and 17.59\% on average, and the average RA reductions are 33.39\% and 30.35\% within 2h and 3h, respectively. 

These results indicate that CM-CASL achieves better CA and RA than the state-of-the-art regression-based and rank-based performance modeling approaches. These improvements are attributed to the exploitation of unlabeled configurations in CM-CASL. In addition, we observe that the experiments with selection ratio 1/3 achieves better CA and RA than with other ratios, which is due to the fact that this setting achieves a better balance between initial labeled samples and manual labels. A small number of initial samples result in a weak initial classifier, and the subsequent selection of promising samples in AL is also affected. On the other hand, a larger initial training dataset means less utilization of tuning experiences, hence restricting further performance improvements of the classifier.

\subsection{Impacts of AL and SSL}\label{impacts}
The above results demonstrate that the integration of AL and SSL contributes significantly to the performance improvement of comparison-based models. However, the magnitude of the effect of AL and SSL on the performance of CM-CASL is still unclear. To gain a deeper insight, we perform ablation experiments to investigate the effects of AL and SSL separately. Specifically, we remove the SSL component in Algorithm~\ref{alg:CM-CASL} and obtain the AL approach, called AL-IR, which takes informativeness and representativeness into consideration to select promising configurations. The AL approach, referred to as AL-I, which only consider the candidates' informativeness, is added as another baseline to demonstrate the effectiveness of our proposed query strategy. If we turn off the AL phase in Algorithm~\ref{alg:CM-CASL}, the AL-based pseudolabel verification does not work, hence yielding a pure SSL algorithm. To assess the impact of different distances in pseudolabel assignment, we slightly modify the original CM-CASL algorithm as ASSL-H, which assigns pseudolabels to unlabeled samples at the farthest distances to the decision boundary in SSL iterations. Moreover, a base learner (i.e., an SVM classifier) is used in this experiment. 

In our experiments, all the algorithms use the same experimental settings as CM-CASL. It is worth mentioning that the total amount of training data for SVM is the sum of the initial sample size in the other algorithms and the sample size added by AL, where the initial samples are consistent with the other algorithms, while the added samples are randomly selected from the dataset. Taking the experimental settings of 2h and selection ratio 1/2 as an example, a summary of performance of SVM, AL-I, AI-IR, SSL, ASSL-H and CM-CASL is presented in Table~\ref{tab:impact}. 
\begin{table}
\caption{Normalized CAs of SVM, AL-I, AI-IR, SSL, ASSL-H and CM-CASL over
different subject systems and workloads. All CAs are normalized to the corresponding CAs of SVM.}
\label{tab:impact}
 \resizebox{\linewidth}{!}{
\begin{tabular}{lcccccc}
\hline
Systems  & SVM & AL-I    & AL-IR    & SSL      & ASSL-H            & CM-CASL          \\
\hline
Hadoop-S & 1 & 1.1064          & 1.0689          & 1.0516 & 1.1158          & \textbf{1.1800} \\
Hadoop-T & 1 & 1.0063          & 0.9514          & 1.1082 & 1.0564          & \textbf{1.1301} \\
Hadoop-W & 1 & 1.0492          & 0.9966          & 1.0390 & \textbf{1.0594} & \textbf{1.0594} \\
Spark-S  & 1 & 0.8866          & 1.0405          & 0.7139 & 1.0526          & \textbf{1.0567} \\
Spark-T  & 1 & 1.2505          & 1.2147          & 1.0716 & 1.2883          & \textbf{1.3002} \\
Spark-W  & 1 & 0.8600          & 1.1549          & 0.7038 & 1.1896          & \textbf{1.2441} \\
MySQL    & 1 & \textbf{1.2650} & 1.2258          & 0.8603 & 1.2533          & 1.2572          \\
SQLite   & 1 & \textbf{1.1304} & 1.0885          & 0.9127 & 1.1257          & 1.1176          \\
Redis    & 1 & 1.0542          & 1.1697          & 0.9110 & 1.1919          & \textbf{1.1975} \\
Tomcat   & 1 & 0.9105          & \textbf{1.1153} & 0.9225 & 1.1050          & 1.0981          \\
x264     & 1 & 1.2401          & 1.2356          & 0.6607 & \textbf{1.2559} & 1.2492          \\
\hline
AVR      & 1 & 1.0690          & 1.1147          & 0.9050 & 1.1540          & 1.1718          \\
VAR      & 0 & 0.0177          & 0.0075          & 0.0206 & 0.0062          & 0.0060           \\
\hline
\end{tabular}}
\end{table}

These results show that CM-CASL performs the best among the six comparison-based modeling approaches in 7 out of 11 test cases, ASSL-H and AL-I perform the best in 2 test cases, and AL-IR performs the best in 1 test case. On average, CM-CASL outperforms all other approaches. It reduces the average CA by 17.18\%, 9.62\%, 5.12\%, 29.48\% and 1.54\% compared with SVM, AL-I, AL-IR, SSL and ASSL-H, respectively. Moreover, it is worth noting that the approaches leveraging AL (AL-I, AL-IR, ASSL-H and CM-CASL) significantly outperform those that do not employ AL (SVM and SSL). The above observations indicate that AL plays a significant role in CM-CASL, and SSL as a supplement to AL further improves the performance and robustness of CM-CASL.

By comparing the CAs of SVM, AL-I, and AL-IR, we can further verify the effect of AL. Both AL approaches achieve better accuracy than SVM, which randomly selects training samples (i.e., a passive learner). Moreover, AL-IR achieves better performance more stably than AI-I, because considering both informativeness and representativeness of unlabeled samples in the query strategy leads to a higher quality of training samples. Consequently, we believe that the adoption of AL-IR in the framework of CM-CASL may facilitate a deep fusion of expert knowledge.

In the SSL approach, the samples with the highest confidence are selected and assigned with the predicted labels by the current classifier without human involvement. However, when the accuracy of the initial classifier is low, it may introduce many incorrect pseudolabels during the training process, and therefore deteriorate the performance of classifier. In addition, samples with the highest confidence may not provide more information for classifier training, so the performance of the SSL approach is weaker than that of the AL-based approaches.

Although it is well understood that SSL might deteriorate performance in some cases, adopting both AL and SSL (ASSL-H and CM-CASL) achieves the best performance most of the time. This is due to the introduction of AL-based pseudolabel verification, which guarantees the credibility of added samples. The higher performance of CM-CASL than ASSL-H further validates that the samples at median distances to the decision boundary would be more informative than the samples with the farthest distances (i.e., the highest confidence), thus helping AL find promising samples more accurately. 

The above observation reveals the impacts of AL and SSL in CM-CASL, and the experimental results show that CM-CASL exhibits more stable performance in comparison with other algorithms (i.e., less performance variance), which is always favored by users.

\subsection{Sensitivity to Expert Accuracy}\label{sensitivity}
Expert accuracy refers to the probability of experts correctly labeling a comparison-based sample in the AL process. Here, we explore the influence of different expert accuracies on algorithm performance (i.e., CA). The experiment conducted here uses the same parameter setting as the previous experiments, and the different expert accuracies are set to be 0.7, 0.8, 0.9, and 1, respectively. The participants with expert accuracies 0.8 and 0.9 are selected from the participants with working years $5\leq T_w<10$, while the participants with expert accuracies 0.7 and 1 are selected from the participants with working years $T_w<5$ and $T_w\geq10$, respectively.
\begin{figure}[h]
  \centering
  \includegraphics[width=\linewidth]{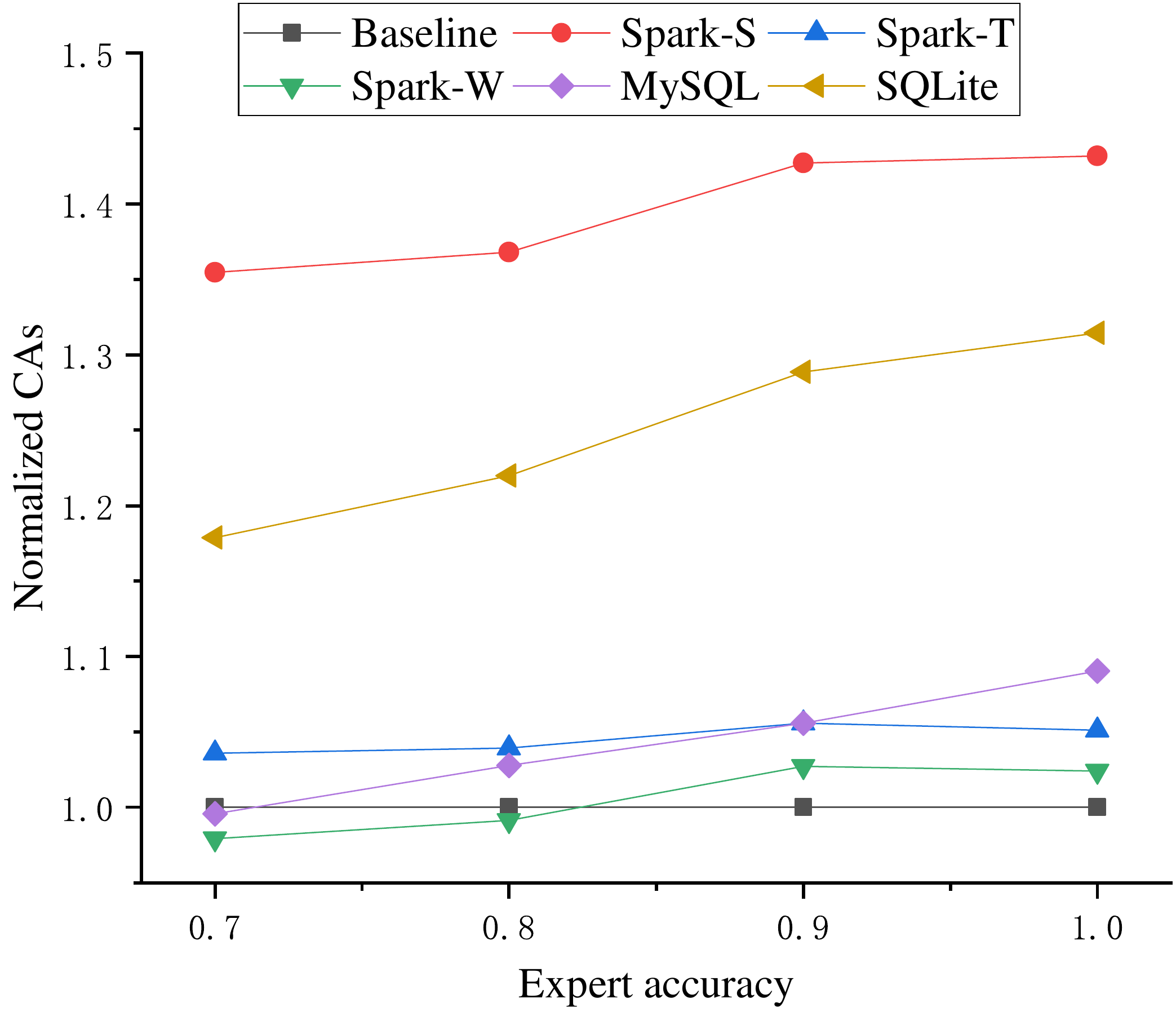}
  \caption{Normalized CAs of baseline and CM-CASL with different expert accuracies (normalized to the corresponding CA of baseline).}
  \label{expert}
\end{figure}

Figure~\ref{expert} compares the normalized CAs of the baseline and CM-CASL with different expert accuracies in five test cases. The baseline represents the better result of DeepPerf and R-CART. As the expert accuracy decreases, the performance of CM-CASL also decreases slightly. Specifically, the performance of CM-CASL in MySQL is surpassed by the baseline when the expert accuracy drops to 0.7, and this value reaches 0.8 with the Wordcount workload in Spark. In other cases, CM-CASL still achieves good results when the expert accuracy is reduced to 0.7, i.e., obtains better CAs than the baseline. These experimental results indicate that CM-CASL is robust to different expert accuracies. To ensure the algorithm performance, we recommend using CM-CASL when the expert accuracy is greater than 0.7, which is easily achieved when the expert is familiar with the subject software system.

\subsection{Improving Configuration Tuning}\label{improving}
We evaluate whether CM-CASL improves the efficacy of configuration tuning by providing a better performance model. We follow the two-phase learning-based configuration tuning process, where a performance model is built in the first phase, and in the second phase, good configurations are recommended by applying a search strategy to the configuration space and providing the performance comparison results by the trained model.

Since the focus of this paper is on the first phase, i.e., performance modeling, we only use genetic algorithm (GA) as the search strategy to evaluate the effect of different performance models on tuning efficacy. GA is a global optimization algorithm which mainly consists of selection, crossover and mutation operations. It is well-known for being robust against local optima, and is suitable for searching for the optimal configuration in a complex configuration space \citep{bei2015rfhoc,hua2018hadoop,bei2017mest,yu2018datasize,tang2017system,trotter2019forecasting}.

For the performance modeling phase, we compare our CM-CASL approach with two state-of-the-art baselines, namely, DeepPerf and R-CART. The whole tuning approaches with the above three performance models are denoted as CM-CASL+GA, DeepPerf+GA, and R-CAST+GA, respectively. As in the experiments above, we set the same time budget (i.e., 2h) for the three different performance modeling approaches. The trained performance models and GA are employed to find promising configurations in the configuration space. In our experiments, GA has a population size of 200, a crossover rate of 0.5, and a mutation rate of 0.015. Finally, tuning efficacy is evaluated by the performance of the tuned configurations.

\begin{figure}[h]
  \centering
  \includegraphics[width=\linewidth]{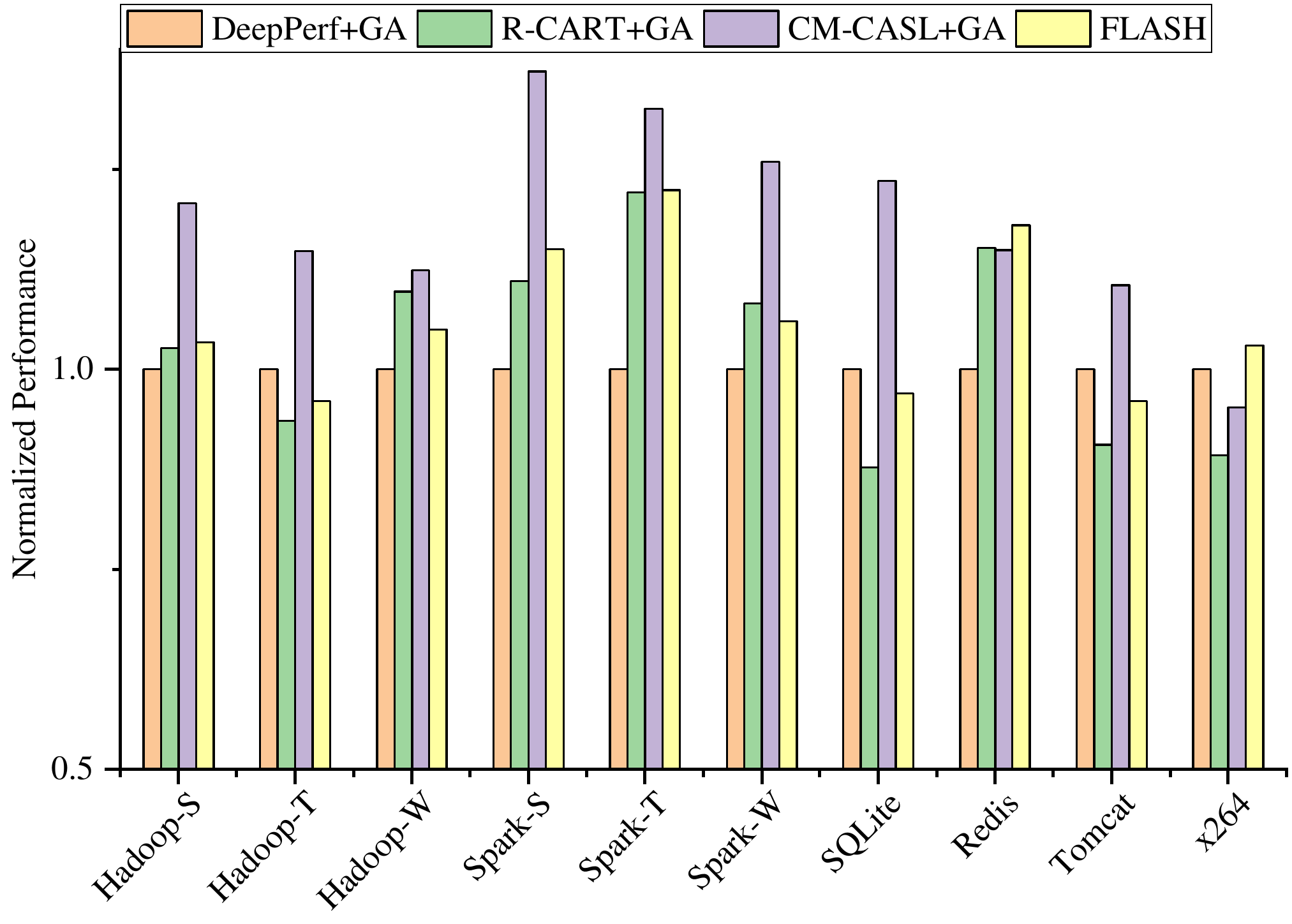}
  \caption{Normalized tuned performance of four tuning approaches (normalized to the corresponding performance of DeepPerf+GA).}
  \label{fig:tuned}
\end{figure}

In addition, we compare our approach with a state-of-the-art sequential model-based tuning method, namely FLASH \citep{nair2018finding}. Following the framework of Sequential Model-based Optimization (SMBO), FLASH builds surrogate model using CART and employs Maximum Mean as acquisition function. That is, FLASH actively selects the configuration with the maximum predicted performance for next evaluation, with the predicted performance provided by the surrogate model. We use the same initial labeled samples and the same tuning time budget for all tuning approaches. The normalized tuned performance of the above four tuning approaches is shown in Figure~\ref{fig:tuned}.

We observe that CM-CASL+GA achieves the best tuned performance in almost all the subject systems and workloads. Specifically, CM-CASL+GA improves the tuned performance by an average of 18.76\%, 16.05\%, and 12.19\% over DeepPerf+GA, R-CAST+GA, and FLASH, respectively. In the comparison of the two-phase tuning approaches, the improvement in tuning efficacy is attributed to the fact that CM-CASL provides a performance model with higher CA and RA, while comparing the performance of different configurations is one fundamental strategy for configuration tuning. 

On the other hand, the surrogate model in FLASH is designed to guide the search process, it can only provide accurate predictions for configurations that lie in the search path of the tuning process, not for the entire configuration space. Meanwhile, the acquisition function of FLASH relies heavily on the performance predictions provided by the surrogate model. This may cause a reduction in search efficiency and hence tuning efficacy in scenarios where tuning time is limited. Moreover, the results validate that performance modeling is critical for the learning-based configuration tuning, as the accuracy of the performance model has a significant impact on the tuning efficacy.

\subsection{Threats to Validity}\label{threats}
\textbf{Internal validity}: To increase the internal validity, we perform controlled experiments by executing each test case for three times and calculating the average of these three runs. Such a method can avoid the misleading effects of specifically selected test cases and ensure the stability of the results. Moreover, we use the same initial labeled samples and the same TCs for all algorithms to compare the CA and RA in each test case. These results are considered to be fair and reliable. In addition, the hyperparameters for our algorithm and two baseline algorithms are set by employing a simple but effective parameter tuning algorithm, grid search.

\textbf{External Validity}: We increase the external validity by choosing seven different SUTs, including two big-data processing systems (Hadoop and Spark), three database systems (MySQL, SQLite and Redis), one video encoder (x264), and one web application server (Tomcat). Furthermore, we choose three HiBench workloads for Hadoop and Spark, namely, Sort, Terasort, and Wordcount. The results of CM-CASL are qualitatively similar and are predicted to be applicable to other software systems.

\section{Related Work}\label{related}
\textbf{Performance Modeling for Software Systems}. Software systems have a large number of configuration parameters that controlling nearly all aspects of runtime operation. Researchers make an effort to understand the relationship of these parameters and system performance. Several performance prediction models \citep{ha2019deepperf, guo2013variability, guo2018data, nair2018faster, sarkar2015cost, valov2015empirical, nair2017using, zhang2015performance, zhang2016mathematical, kolesnikov2019tradeoffs} have been proposed to explore that relationship. These approaches apply regression techniques to model the relationship. The most commonly used models include CART \citep{guo2013variability, guo2018data, nair2018faster, sarkar2015cost, valov2015empirical, nair2017using}, RF \citep{valov2015empirical, bei2015rfhoc}, neural networks \citep{ha2019deepperf, mahgoub2017rafiki, zheng2014self}, Gaussian process regression \citep{duan2009tuning, thummala2010ituned, van2017automatic, zhang2018demonstration}, and Support Vector Regression (SVR) \citep{chen2019regression, valov2015empirical}.

However, a regression-based performance model with high prediction accuracy may not be able to accurately predict which of two configurations performs better \citep{chen2019regression}. Comparison-based performance models \citep{bao2018autoconfig, zhu2019classytune} and rank-based performance models \citep{nair2017using} are introduced to solve this problem.

Comparison-based approaches \citep{bao2018autoconfig, zhu2019classytune} model the performance comparison relation between a configuration pair. The developed models can be leveraged to search for the optimal configuration using various algorithms such as a multiple bound-and-search algorithm \citep{bao2018autoconfig}, a clustering-based tuning algorithm \citep{zhu2019classytune}, etc. Rank-based approaches \citep{nair2017using} treat the original regression problem as a ranking problem, i.e., ranking configurations based on their predicted performance and choosing good configurations rather than accurately predicting performance for all configurations.

\textbf{AL and SSL for Performance Modeling}. In many real-world applications, obtaining labeled samples is very costly, while a large number of unlabeled samples are readily available. To exploit unlabeled samples and improve the accuracies of learners, AL and SSL have been extensively investigated for many real-world problems in machine learning, such as text classification \citep{tong2001support, hoi2006large, burkhardt2018semisupervised}, information extraction \citep{thompson1999active, wu2005semi}, image classification and retrieval \citep{hoi2005semi, li2013combining, pedronette2019semi}, and cancer diagnosis \citep{nguyen2020active, menon2020deep}. 

AL improves the prediction accuracy by querying the oracle for the labels of some unlabeled samples \citep{tong2001active,lewis1994heterogeneous,dasgupta2008hierarchical,huang2010active}. SSL is another mainstream methodology for exploiting unlabeled data to improve prediction accuracy. Unlike AL, SSL aims to label samples by the learner itself, where no human intervention is assumed \citep{shahshahani1994effect,zhou2010semi,miller1996mixture,bennett1998semi,chapelle2008optimization,zhu2003semi,zhou2007semisupervised, zhou2007semi}.

However, there are few studies using AL and SSL in performance modeling of software systems. ClassyTune \citep{zhu2019classytune} adopts a comparison-based model, and introduces manual tuning experiences by generating comparison-based samples to augment the training data. But there is no discussion about the query strategy in ClassyTune. Moreover, some studies for performance prediction add more samples during the performance modeling of software systems \citep{guo2018data, sarkar2015cost, chen2019regression}. Additional samples are needed when the resulting performance prediction model does not meet the performance requirements. Acquiring additional samples usually follows a feature-size heuristic, i.e., randomly sampling some configurations with their performance. This method is helpful for reducing the cost of performance measurements, but it fails to effectively use unlabeled samples to improve model performance.

\section{Conclusion}\label{conclusion}
In this paper, we proposed a novel approach, CM-CASL, for comparison-based performance modeling of software systems. By combining AL and SSL in a collaborative manner, CM-CASL is able to acquire more high-quality training samples without additional running cost to improve the performance of the final classifier. To demonstrate the efficacy of CM-CASL, we conducted comprehensive experiments in diverse software systems. Experimental results show that, given the same modeling time constraint, CM-CASL outperforms two state-of-the-art models by 17.59\%-28.97\% in terms of classification accuracy and 30.11\%-35.38\% in terms of rank accuracy, on average. Furthermore, we conducted experiments to demonstrate the performance superiority of CM-CASL in terms of prediction accuracy over the base learner, pure SSL, pure AL with different query strategies, and a slightly modified CM-CASL. Finally, we verified that CM-CASL can improve the tuning efficacy of the learning-based configuration tuning by providing a better performance model.

\section*{Acknowledgments}
This work is supported by the National Natural Science Foundation of China [Grant No. 62172316]; the Ministry of Education Humanities and Social Science Project of China [Grant No. 17YJA790047]; the Soft Science Research Plans of Shaanxi Province [Grant No. 2020KRZ018]; the Research Project on Major Theoretical and Practical Problems of Philosophy and Social Sciences in Shaanxi Province [Grant No. 20JZ-25]; the Key R\&D Program of Shaanxi [Grant No. 2019ZDLGY13-03-02]; the Natural Science Foundation of Shaanxi Province, China [Grant No. 2019JM-368]; and the Key R\&D Program of Hebei [Grant No. 20310102D].

\bibliographystyle{elsarticle-harv} 
\bibliography{mybibfile}

\begin{thebibliography}{67}
\expandafter\ifx\csname natexlab\endcsname\relax\def\natexlab#1{#1}\fi
\providecommand{\url}[1]{\texttt{#1}}
\providecommand{\href}[2]{#2}
\providecommand{\path}[1]{#1}
\providecommand{\DOIprefix}{doi:}
\providecommand{\ArXivprefix}{arXiv:}
\providecommand{\URLprefix}{URL: }
\providecommand{\Pubmedprefix}{pmid:}
\providecommand{\doi}[1]{\href{http://dx.doi.org/#1}{\path{#1}}}
\providecommand{\Pubmed}[1]{\href{pmid:#1}{\path{#1}}}
\providecommand{\bibinfo}[2]{#2}
\ifx\xfnm\relax \def\xfnm[#1]{\unskip,\space#1}\fi
\bibitem[{Babu et~al.(2009)Babu, Borisov, Duan, Herodotou and
  Thummala}]{babu2009automated}
\bibinfo{author}{Babu, S.}, \bibinfo{author}{Borisov, N.},
  \bibinfo{author}{Duan, S.}, \bibinfo{author}{Herodotou, H.},
  \bibinfo{author}{Thummala, V.}, \bibinfo{year}{2009}.
\newblock \bibinfo{title}{Automated experiment-driven management of (database)
  systems.}, in: \bibinfo{booktitle}{HotOS}.
\bibitem[{Bao et~al.(2018a)Bao, Liu and Chen}]{bao2018learning}
\bibinfo{author}{Bao, L.}, \bibinfo{author}{Liu, X.}, \bibinfo{author}{Chen,
  W.}, \bibinfo{year}{2018}a.
\newblock \bibinfo{title}{Learning-based automatic parameter tuning for big
  data analytics frameworks}, in: \bibinfo{booktitle}{2018 IEEE International
  Conference on Big Data (Big Data)}, \bibinfo{organization}{IEEE}. pp.
  \bibinfo{pages}{181--190}.
\bibitem[{Bao et~al.(2019)Bao, Liu, Wang and Fang}]{bao2019actgan}
\bibinfo{author}{Bao, L.}, \bibinfo{author}{Liu, X.}, \bibinfo{author}{Wang,
  F.}, \bibinfo{author}{Fang, B.}, \bibinfo{year}{2019}.
\newblock \bibinfo{title}{Actgan: Automatic configuration tuning for software
  systems with generative adversarial networks}, in: \bibinfo{booktitle}{2019
  34th IEEE/ACM International Conference on Automated Software Engineering
  (ASE)}, \bibinfo{organization}{IEEE}. pp. \bibinfo{pages}{465--476}.
\bibitem[{Bao et~al.(2018b)Bao, Liu, Xu and Fang}]{bao2018autoconfig}
\bibinfo{author}{Bao, L.}, \bibinfo{author}{Liu, X.}, \bibinfo{author}{Xu, Z.},
  \bibinfo{author}{Fang, B.}, \bibinfo{year}{2018}b.
\newblock \bibinfo{title}{Autoconfig: Automatic configuration tuning for
  distributed message systems}, in: \bibinfo{booktitle}{2018 33rd IEEE/ACM
  International Conference on Automated Software Engineering (ASE)},
  \bibinfo{organization}{IEEE}. pp. \bibinfo{pages}{29--40}.
\bibitem[{Bei et~al.(2017)Bei, Yu, Liu, Xu, Feng and Song}]{bei2017mest}
\bibinfo{author}{Bei, Z.}, \bibinfo{author}{Yu, Z.}, \bibinfo{author}{Liu, Q.},
  \bibinfo{author}{Xu, C.}, \bibinfo{author}{Feng, S.}, \bibinfo{author}{Song,
  S.}, \bibinfo{year}{2017}.
\newblock \bibinfo{title}{Mest: A model-driven efficient searching approach for
  mapreduce self-tuning}.
\newblock \bibinfo{journal}{IEEE Access} \bibinfo{volume}{5},
  \bibinfo{pages}{3580--3593}.
\bibitem[{Bei et~al.(2015)Bei, Yu, Zhang, Xiong, Xu, Eeckhout and
  Feng}]{bei2015rfhoc}
\bibinfo{author}{Bei, Z.}, \bibinfo{author}{Yu, Z.}, \bibinfo{author}{Zhang,
  H.}, \bibinfo{author}{Xiong, W.}, \bibinfo{author}{Xu, C.},
  \bibinfo{author}{Eeckhout, L.}, \bibinfo{author}{Feng, S.},
  \bibinfo{year}{2015}.
\newblock \bibinfo{title}{Rfhoc: A random-forest approach to auto-tuning
  hadoop's configuration}.
\newblock \bibinfo{journal}{IEEE Transactions on Parallel and Distributed
  Systems} \bibinfo{volume}{27}, \bibinfo{pages}{1470--1483}.
\bibitem[{Bennett and Demiriz(1998)}]{bennett1998semi}
\bibinfo{author}{Bennett, K.}, \bibinfo{author}{Demiriz, A.},
  \bibinfo{year}{1998}.
\newblock \bibinfo{title}{Semi-supervised support vector machines}.
\newblock \bibinfo{journal}{Advances in Neural Information processing systems}
  \bibinfo{volume}{11}.
\bibitem[{Blum and Mitchell(1998)}]{blum1998combining}
\bibinfo{author}{Blum, A.}, \bibinfo{author}{Mitchell, T.},
  \bibinfo{year}{1998}.
\newblock \bibinfo{title}{Combining labeled and unlabeled data with
  co-training}, in: \bibinfo{booktitle}{Proceedings of the eleventh annual
  conference on Computational learning theory}, pp. \bibinfo{pages}{92--100}.
\bibitem[{Bruzzone and Persello(2010)}]{bruzzone2010recent}
\bibinfo{author}{Bruzzone, L.}, \bibinfo{author}{Persello, C.},
  \bibinfo{year}{2010}.
\newblock \bibinfo{title}{Recent trends in classification of remote sensing
  data: Active and semisupervised machine learning paradigms}, in:
  \bibinfo{booktitle}{2010 IEEE International Geoscience and Remote Sensing
  Symposium}, \bibinfo{organization}{IEEE}. pp. \bibinfo{pages}{3720--3723}.
\bibitem[{Burkhardt et~al.(2018)Burkhardt, Siekiera and
  Kramer}]{burkhardt2018semisupervised}
\bibinfo{author}{Burkhardt, S.}, \bibinfo{author}{Siekiera, J.},
  \bibinfo{author}{Kramer, S.}, \bibinfo{year}{2018}.
\newblock \bibinfo{title}{Semisupervised bayesian active learning for text
  classification}, in: \bibinfo{booktitle}{Bayesian Deep Learning Workshop at
  NeurIPS}.
\bibitem[{Chapelle et~al.(2008)Chapelle, Sindhwani and
  Keerthi}]{chapelle2008optimization}
\bibinfo{author}{Chapelle, O.}, \bibinfo{author}{Sindhwani, V.},
  \bibinfo{author}{Keerthi, S.S.}, \bibinfo{year}{2008}.
\newblock \bibinfo{title}{Optimization techniques for semi-supervised support
  vector machines.}
\newblock \bibinfo{journal}{Journal of Machine Learning Research}
  \bibinfo{volume}{9}.
\bibitem[{Chen et~al.(2015)Chen, Zhuo, Yeh, Lin and Liao}]{chen2015machine}
\bibinfo{author}{Chen, C.O.}, \bibinfo{author}{Zhuo, Y.Q.},
  \bibinfo{author}{Yeh, C.C.}, \bibinfo{author}{Lin, C.M.},
  \bibinfo{author}{Liao, S.W.}, \bibinfo{year}{2015}.
\newblock \bibinfo{title}{Machine learning-based configuration parameter tuning
  on hadoop system}, in: \bibinfo{booktitle}{2015 IEEE International Congress
  on Big Data}, \bibinfo{organization}{IEEE}. pp. \bibinfo{pages}{386--392}.
\bibitem[{Chen et~al.(2019)Chen, Gu, He and Xuan}]{chen2019regression}
\bibinfo{author}{Chen, Y.}, \bibinfo{author}{Gu, Y.}, \bibinfo{author}{He, L.},
  \bibinfo{author}{Xuan, J.}, \bibinfo{year}{2019}.
\newblock \bibinfo{title}{Regression models for performance ranking of
  configurable systems: A comparative study}, in:
  \bibinfo{booktitle}{International Workshop on Structured Object-Oriented
  Formal Language and Method}, \bibinfo{organization}{Springer}. pp.
  \bibinfo{pages}{243--258}.
\bibitem[{Dasgupta(2011)}]{dasgupta2011two}
\bibinfo{author}{Dasgupta, S.}, \bibinfo{year}{2011}.
\newblock \bibinfo{title}{Two faces of active learning}.
\newblock \bibinfo{journal}{Theoretical computer science}
  \bibinfo{volume}{412}, \bibinfo{pages}{1767--1781}.
\bibitem[{Dasgupta and Hsu(2008)}]{dasgupta2008hierarchical}
\bibinfo{author}{Dasgupta, S.}, \bibinfo{author}{Hsu, D.},
  \bibinfo{year}{2008}.
\newblock \bibinfo{title}{Hierarchical sampling for active learning}, in:
  \bibinfo{booktitle}{Proceedings of the 25th international conference on
  Machine learning}, pp. \bibinfo{pages}{208--215}.
\bibitem[{Duan et~al.(2009)Duan, Thummala and Babu}]{duan2009tuning}
\bibinfo{author}{Duan, S.}, \bibinfo{author}{Thummala, V.},
  \bibinfo{author}{Babu, S.}, \bibinfo{year}{2009}.
\newblock \bibinfo{title}{Tuning database configuration parameters with
  ituned}.
\newblock \bibinfo{journal}{Proceedings of the VLDB Endowment}
  \bibinfo{volume}{2}, \bibinfo{pages}{1246--1257}.
\bibitem[{Guo et~al.(2013)Guo, Czarnecki, Apel, Siegmund and
  Wasowski}]{guo2013variability}
\bibinfo{author}{Guo, J.}, \bibinfo{author}{Czarnecki, K.},
  \bibinfo{author}{Apel, S.}, \bibinfo{author}{Siegmund, N.},
  \bibinfo{author}{Wasowski, A.}, \bibinfo{year}{2013}.
\newblock \bibinfo{title}{Variability-aware performance prediction: A
  statistical learning approach}, in: \bibinfo{booktitle}{2013 28th IEEE/ACM
  International Conference on Automated Software Engineering (ASE)},
  \bibinfo{organization}{IEEE}. pp. \bibinfo{pages}{301--311}.
\bibitem[{Guo et~al.(2018)Guo, Yang, Siegmund, Apel, Sarkar, Valov, Czarnecki,
  Wasowski and Yu}]{guo2018data}
\bibinfo{author}{Guo, J.}, \bibinfo{author}{Yang, D.},
  \bibinfo{author}{Siegmund, N.}, \bibinfo{author}{Apel, S.},
  \bibinfo{author}{Sarkar, A.}, \bibinfo{author}{Valov, P.},
  \bibinfo{author}{Czarnecki, K.}, \bibinfo{author}{Wasowski, A.},
  \bibinfo{author}{Yu, H.}, \bibinfo{year}{2018}.
\newblock \bibinfo{title}{Data-efficient performance learning for configurable
  systems}.
\newblock \bibinfo{journal}{Empirical Software Engineering}
  \bibinfo{volume}{23}, \bibinfo{pages}{1826--1867}.
\bibitem[{Ha and Zhang(2019)}]{ha2019deepperf}
\bibinfo{author}{Ha, H.}, \bibinfo{author}{Zhang, H.}, \bibinfo{year}{2019}.
\newblock \bibinfo{title}{Deepperf: performance prediction for configurable
  software with deep sparse neural network}, in: \bibinfo{booktitle}{2019
  IEEE/ACM 41st International Conference on Software Engineering (ICSE)},
  \bibinfo{organization}{IEEE}. pp. \bibinfo{pages}{1095--1106}.
\bibitem[{Hoi et~al.(2006)Hoi, Jin and Lyu}]{hoi2006large}
\bibinfo{author}{Hoi, S.C.}, \bibinfo{author}{Jin, R.}, \bibinfo{author}{Lyu,
  M.R.}, \bibinfo{year}{2006}.
\newblock \bibinfo{title}{Large-scale text categorization by batch mode active
  learning}, in: \bibinfo{booktitle}{Proceedings of the 15th international
  conference on World Wide Web}, pp. \bibinfo{pages}{633--642}.
\bibitem[{Hoi and Lyu(2005)}]{hoi2005semi}
\bibinfo{author}{Hoi, S.C.}, \bibinfo{author}{Lyu, M.R.}, \bibinfo{year}{2005}.
\newblock \bibinfo{title}{A semi-supervised active learning framework for image
  retrieval}, in: \bibinfo{booktitle}{2005 IEEE Computer Society Conference on
  Computer Vision and Pattern Recognition (CVPR'05)},
  \bibinfo{organization}{IEEE}. pp. \bibinfo{pages}{302--309}.
\bibitem[{Hua et~al.(2018)Hua, Huang and Liu}]{hua2018hadoop}
\bibinfo{author}{Hua, X.}, \bibinfo{author}{Huang, M.C.}, \bibinfo{author}{Liu,
  P.}, \bibinfo{year}{2018}.
\newblock \bibinfo{title}{Hadoop configuration tuning with ensemble modeling
  and metaheuristic optimization}.
\newblock \bibinfo{journal}{IEEE Access} \bibinfo{volume}{6},
  \bibinfo{pages}{44161--44174}.
\bibitem[{Huang et~al.(2010a)Huang, Huang, Dai, Xie and
  Huang}]{huang2010hibench}
\bibinfo{author}{Huang, S.}, \bibinfo{author}{Huang, J.}, \bibinfo{author}{Dai,
  J.}, \bibinfo{author}{Xie, T.}, \bibinfo{author}{Huang, B.},
  \bibinfo{year}{2010}a.
\newblock \bibinfo{title}{The hibench benchmark suite: Characterization of the
  mapreduce-based data analysis}, in: \bibinfo{booktitle}{2010 IEEE 26th
  International Conference on Data Engineering Workshops (ICDEW 2010)},
  \bibinfo{organization}{IEEE}. pp. \bibinfo{pages}{41--51}.
\bibitem[{Huang et~al.(2010b)Huang, Jin and Zhou}]{huang2010active}
\bibinfo{author}{Huang, S.J.}, \bibinfo{author}{Jin, R.},
  \bibinfo{author}{Zhou, Z.H.}, \bibinfo{year}{2010}b.
\newblock \bibinfo{title}{Active learning by querying informative and
  representative examples}.
\newblock \bibinfo{journal}{Advances in neural information processing systems}
  \bibinfo{volume}{23}.
\bibitem[{Kolesnikov et~al.(2019)Kolesnikov, Siegmund, K{\"a}stner, Grebhahn
  and Apel}]{kolesnikov2019tradeoffs}
\bibinfo{author}{Kolesnikov, S.}, \bibinfo{author}{Siegmund, N.},
  \bibinfo{author}{K{\"a}stner, C.}, \bibinfo{author}{Grebhahn, A.},
  \bibinfo{author}{Apel, S.}, \bibinfo{year}{2019}.
\newblock \bibinfo{title}{Tradeoffs in modeling performance of highly
  configurable software systems}.
\newblock \bibinfo{journal}{Software \& Systems Modeling} \bibinfo{volume}{18},
  \bibinfo{pages}{2265--2283}.
\bibitem[{Kremer et~al.(2014)Kremer, Steenstrup~Pedersen and
  Igel}]{kremer2014active}
\bibinfo{author}{Kremer, J.}, \bibinfo{author}{Steenstrup~Pedersen, K.},
  \bibinfo{author}{Igel, C.}, \bibinfo{year}{2014}.
\newblock \bibinfo{title}{Active learning with support vector machines}.
\newblock \bibinfo{journal}{Wiley Interdisciplinary Reviews: Data Mining and
  Knowledge Discovery} \bibinfo{volume}{4}, \bibinfo{pages}{313--326}.
\bibitem[{Lewis and Catlett(1994)}]{lewis1994heterogeneous}
\bibinfo{author}{Lewis, D.D.}, \bibinfo{author}{Catlett, J.},
  \bibinfo{year}{1994}.
\newblock \bibinfo{title}{Heterogeneous uncertainty sampling for supervised
  learning}, in: \bibinfo{booktitle}{Machine learning proceedings 1994}.
  \bibinfo{publisher}{Elsevier}, pp. \bibinfo{pages}{148--156}.
\bibitem[{Lewis and Gale(1994)}]{lewis1994sequential}
\bibinfo{author}{Lewis, D.D.}, \bibinfo{author}{Gale, W.A.},
  \bibinfo{year}{1994}.
\newblock \bibinfo{title}{A sequential algorithm for training text
  classifiers}, in: \bibinfo{booktitle}{SIGIR’94},
  \bibinfo{organization}{Springer}. pp. \bibinfo{pages}{3--12}.
\bibitem[{Li et~al.(2013)Li, Wang and Tang}]{li2013combining}
\bibinfo{author}{Li, M.}, \bibinfo{author}{Wang, R.}, \bibinfo{author}{Tang,
  K.}, \bibinfo{year}{2013}.
\newblock \bibinfo{title}{Combining semi-supervised and active learning for
  hyperspectral image classification}, in: \bibinfo{booktitle}{2013 IEEE
  Symposium on Computational Intelligence and Data Mining (CIDM)},
  \bibinfo{organization}{IEEE}. pp. \bibinfo{pages}{89--94}.
\bibitem[{Mahgoub et~al.(2017)Mahgoub, Wood, Ganesh, Mitra, Gerlach, Harrison,
  Meyer, Grama, Bagchi and Chaterji}]{mahgoub2017rafiki}
\bibinfo{author}{Mahgoub, A.}, \bibinfo{author}{Wood, P.},
  \bibinfo{author}{Ganesh, S.}, \bibinfo{author}{Mitra, S.},
  \bibinfo{author}{Gerlach, W.}, \bibinfo{author}{Harrison, T.},
  \bibinfo{author}{Meyer, F.}, \bibinfo{author}{Grama, A.},
  \bibinfo{author}{Bagchi, S.}, \bibinfo{author}{Chaterji, S.},
  \bibinfo{year}{2017}.
\newblock \bibinfo{title}{Rafiki: A middleware for parameter tuning of nosql
  datastores for dynamic metagenomics workloads}, in:
  \bibinfo{booktitle}{Proceedings of the 18th ACM/IFIP/USENIX Middleware
  Conference}, pp. \bibinfo{pages}{28--40}.
\bibitem[{Menon et~al.(2020)Menon, Chapman, Nguyen, Yesha, Morris and
  Saboury}]{menon2020deep}
\bibinfo{author}{Menon, S.}, \bibinfo{author}{Chapman, D.},
  \bibinfo{author}{Nguyen, P.}, \bibinfo{author}{Yesha, Y.},
  \bibinfo{author}{Morris, M.}, \bibinfo{author}{Saboury, B.},
  \bibinfo{year}{2020}.
\newblock \bibinfo{title}{Deep expectation-maximization for semi-supervised
  lung cancer screening}.
\newblock \bibinfo{journal}{arXiv preprint arXiv:2010.01173} .
\bibitem[{Miller and Uyar(1996)}]{miller1996mixture}
\bibinfo{author}{Miller, D.J.}, \bibinfo{author}{Uyar, H.},
  \bibinfo{year}{1996}.
\newblock \bibinfo{title}{A mixture of experts classifier with learning based
  on both labelled and unlabelled data}.
\newblock \bibinfo{journal}{Advances in neural information processing systems}
  \bibinfo{volume}{9}.
\bibitem[{Nair et~al.(2017)Nair, Menzies, Siegmund and Apel}]{nair2017using}
\bibinfo{author}{Nair, V.}, \bibinfo{author}{Menzies, T.},
  \bibinfo{author}{Siegmund, N.}, \bibinfo{author}{Apel, S.},
  \bibinfo{year}{2017}.
\newblock \bibinfo{title}{Using bad learners to find good configurations}, in:
  \bibinfo{booktitle}{Proceedings of the 2017 11th Joint Meeting on Foundations
  of Software Engineering}, pp. \bibinfo{pages}{257--267}.
\bibitem[{Nair et~al.(2018a)Nair, Menzies, Siegmund and Apel}]{nair2018faster}
\bibinfo{author}{Nair, V.}, \bibinfo{author}{Menzies, T.},
  \bibinfo{author}{Siegmund, N.}, \bibinfo{author}{Apel, S.},
  \bibinfo{year}{2018}a.
\newblock \bibinfo{title}{Faster discovery of faster system configurations with
  spectral learning}.
\newblock \bibinfo{journal}{Automated Software Engineering}
  \bibinfo{volume}{25}, \bibinfo{pages}{247--277}.
\bibitem[{Nair et~al.(2018b)Nair, Yu, Menzies, Siegmund and
  Apel}]{nair2018finding}
\bibinfo{author}{Nair, V.}, \bibinfo{author}{Yu, Z.}, \bibinfo{author}{Menzies,
  T.}, \bibinfo{author}{Siegmund, N.}, \bibinfo{author}{Apel, S.},
  \bibinfo{year}{2018}b.
\newblock \bibinfo{title}{Finding faster configurations using flash}.
\newblock \bibinfo{journal}{IEEE Transactions on Software Engineering}
  \bibinfo{volume}{46}, \bibinfo{pages}{794--811}.
\bibitem[{Nguyen et~al.(2020)Nguyen, Chapman, Menon, Morris and
  Yesha}]{nguyen2020active}
\bibinfo{author}{Nguyen, P.}, \bibinfo{author}{Chapman, D.},
  \bibinfo{author}{Menon, S.}, \bibinfo{author}{Morris, M.},
  \bibinfo{author}{Yesha, Y.}, \bibinfo{year}{2020}.
\newblock \bibinfo{title}{Active semi-supervised expectation maximization
  learning for lung cancer detection from computerized tomography (ct) images
  with minimally label training data}, in: \bibinfo{booktitle}{Medical Imaging
  2020: Computer-Aided Diagnosis}, \bibinfo{organization}{International Society
  for Optics and Photonics}. p. \bibinfo{pages}{113142E}.
\bibitem[{Novakovi{\'c} et~al.(2017)Novakovi{\'c}, Veljovi{\'c}, Ili{\'c},
  Papi{\'c} and Milica}]{novakovic2017evaluation}
\bibinfo{author}{Novakovi{\'c}, J.D.}, \bibinfo{author}{Veljovi{\'c}, A.},
  \bibinfo{author}{Ili{\'c}, S.S.}, \bibinfo{author}{Papi{\'c}, {\v{Z}}.},
  \bibinfo{author}{Milica, T.}, \bibinfo{year}{2017}.
\newblock \bibinfo{title}{Evaluation of classification models in machine
  learning}.
\newblock \bibinfo{journal}{Theory and Applications of Mathematics \& Computer
  Science} \bibinfo{volume}{7}, \bibinfo{pages}{39--46}.
\bibitem[{Pedronette et~al.(2019)Pedronette, Weng, Baldassin and
  Hou}]{pedronette2019semi}
\bibinfo{author}{Pedronette, D.C.G.}, \bibinfo{author}{Weng, Y.},
  \bibinfo{author}{Baldassin, A.}, \bibinfo{author}{Hou, C.},
  \bibinfo{year}{2019}.
\newblock \bibinfo{title}{Semi-supervised and active learning through manifold
  reciprocal knn graph for image retrieval}.
\newblock \bibinfo{journal}{Neurocomputing} \bibinfo{volume}{340},
  \bibinfo{pages}{19--31}.
\bibitem[{Rajan et~al.(2008)Rajan, Ghosh and Crawford}]{rajan2008active}
\bibinfo{author}{Rajan, S.}, \bibinfo{author}{Ghosh, J.},
  \bibinfo{author}{Crawford, M.M.}, \bibinfo{year}{2008}.
\newblock \bibinfo{title}{An active learning approach to hyperspectral data
  classification}.
\newblock \bibinfo{journal}{IEEE Transactions on Geoscience and Remote Sensing}
  \bibinfo{volume}{46}, \bibinfo{pages}{1231--1242}.
\bibitem[{Sarkar et~al.(2015)Sarkar, Guo, Siegmund, Apel and
  Czarnecki}]{sarkar2015cost}
\bibinfo{author}{Sarkar, A.}, \bibinfo{author}{Guo, J.},
  \bibinfo{author}{Siegmund, N.}, \bibinfo{author}{Apel, S.},
  \bibinfo{author}{Czarnecki, K.}, \bibinfo{year}{2015}.
\newblock \bibinfo{title}{Cost-efficient sampling for performance prediction of
  configurable systems (t)}, in: \bibinfo{booktitle}{2015 30th IEEE/ACM
  International Conference on Automated Software Engineering (ASE)},
  \bibinfo{organization}{IEEE}. pp. \bibinfo{pages}{342--352}.
\bibitem[{Settles(2009)}]{settles2009active}
\bibinfo{author}{Settles, B.}, \bibinfo{year}{2009}.
\newblock \bibinfo{title}{Active learning literature survey} .
\bibitem[{Shahshahani and Landgrebe(1994)}]{shahshahani1994effect}
\bibinfo{author}{Shahshahani, B.M.}, \bibinfo{author}{Landgrebe, D.A.},
  \bibinfo{year}{1994}.
\newblock \bibinfo{title}{The effect of unlabeled samples in reducing the small
  sample size problem and mitigating the hughes phenomenon}.
\newblock \bibinfo{journal}{IEEE Transactions on Geoscience and remote sensing}
  \bibinfo{volume}{32}, \bibinfo{pages}{1087--1095}.
\bibitem[{Tang(2017)}]{tang2017system}
\bibinfo{author}{Tang, C.}, \bibinfo{year}{2017}.
\newblock \bibinfo{title}{System performance optimization via design and
  configuration space exploration}, in: \bibinfo{booktitle}{Proceedings of the
  2017 11th Joint Meeting on Foundations of Software Engineering}, pp.
  \bibinfo{pages}{1046--1049}.
\bibitem[{Thompson et~al.(1999)Thompson, Califf and
  Mooney}]{thompson1999active}
\bibinfo{author}{Thompson, C.A.}, \bibinfo{author}{Califf, M.E.},
  \bibinfo{author}{Mooney, R.J.}, \bibinfo{year}{1999}.
\newblock \bibinfo{title}{Active learning for natural language parsing and
  information extraction}, in: \bibinfo{booktitle}{ICML},
  \bibinfo{organization}{Citeseer}. pp. \bibinfo{pages}{406--414}.
\bibitem[{Thummala and Babu(2010)}]{thummala2010ituned}
\bibinfo{author}{Thummala, V.}, \bibinfo{author}{Babu, S.},
  \bibinfo{year}{2010}.
\newblock \bibinfo{title}{ituned: a tool for configuring and visualizing
  database parameters}, in: \bibinfo{booktitle}{Proceedings of the 2010 ACM
  SIGMOD International Conference on Management of data}, pp.
  \bibinfo{pages}{1231--1234}.
\bibitem[{Tong(2001)}]{tong2001active}
\bibinfo{author}{Tong, S.}, \bibinfo{year}{2001}.
\newblock \bibinfo{title}{Active learning: theory and applications}.
\newblock \bibinfo{publisher}{Stanford University}.
\bibitem[{Tong and Chang(2001)}]{tong2001support}
\bibinfo{author}{Tong, S.}, \bibinfo{author}{Chang, E.}, \bibinfo{year}{2001}.
\newblock \bibinfo{title}{Support vector machine active learning for image
  retrieval}, in: \bibinfo{booktitle}{Proceedings of the ninth ACM
  international conference on Multimedia}, pp. \bibinfo{pages}{107--118}.
\bibitem[{Trotter et~al.(2019)Trotter, Wood and Hwang}]{trotter2019forecasting}
\bibinfo{author}{Trotter, M.}, \bibinfo{author}{Wood, T.},
  \bibinfo{author}{Hwang, J.}, \bibinfo{year}{2019}.
\newblock \bibinfo{title}{Forecasting a storm: Divining optimal configurations
  using genetic algorithms and supervised learning}, in:
  \bibinfo{booktitle}{2019 IEEE international conference on autonomic computing
  (ICAC)}, \bibinfo{organization}{IEEE}. pp. \bibinfo{pages}{136--146}.
\bibitem[{Valov et~al.(2015)Valov, Guo and Czarnecki}]{valov2015empirical}
\bibinfo{author}{Valov, P.}, \bibinfo{author}{Guo, J.},
  \bibinfo{author}{Czarnecki, K.}, \bibinfo{year}{2015}.
\newblock \bibinfo{title}{Empirical comparison of regression methods for
  variability-aware performance prediction}, in:
  \bibinfo{booktitle}{Proceedings of the 19th International Conference on
  Software Product Line}, pp. \bibinfo{pages}{186--190}.
\bibitem[{Van~Aken et~al.(2017)Van~Aken, Pavlo, Gordon and
  Zhang}]{van2017automatic}
\bibinfo{author}{Van~Aken, D.}, \bibinfo{author}{Pavlo, A.},
  \bibinfo{author}{Gordon, G.J.}, \bibinfo{author}{Zhang, B.},
  \bibinfo{year}{2017}.
\newblock \bibinfo{title}{Automatic database management system tuning through
  large-scale machine learning}, in: \bibinfo{booktitle}{Proceedings of the
  2017 ACM International Conference on Management of Data}, pp.
  \bibinfo{pages}{1009--1024}.
\bibitem[{Wang et~al.(2016)Wang, Xu and He}]{wang2016novel}
\bibinfo{author}{Wang, G.}, \bibinfo{author}{Xu, J.}, \bibinfo{author}{He, B.},
  \bibinfo{year}{2016}.
\newblock \bibinfo{title}{A novel method for tuning configuration parameters of
  spark based on machine learning}, in: \bibinfo{booktitle}{2016 IEEE 18th
  International Conference on High Performance Computing and Communications;
  IEEE 14th International Conference on Smart City; IEEE 2nd International
  Conference on Data Science and Systems (HPCC/SmartCity/DSS)},
  \bibinfo{organization}{IEEE}. pp. \bibinfo{pages}{586--593}.
\bibitem[{Wu and Pottenger(2005)}]{wu2005semi}
\bibinfo{author}{Wu, T.}, \bibinfo{author}{Pottenger, W.M.},
  \bibinfo{year}{2005}.
\newblock \bibinfo{title}{A semi-supervised active learning algorithm for
  information extraction from textual data}.
\newblock \bibinfo{journal}{Journal of the American Society for Information
  Science and Technology} \bibinfo{volume}{56}, \bibinfo{pages}{258--271}.
\bibitem[{Xu et~al.(2015)Xu, Jin, Fan, Zhou, Pasupathy and
  Talwadker}]{Xu2015Hey}
\bibinfo{author}{Xu, T.}, \bibinfo{author}{Jin, L.}, \bibinfo{author}{Fan, X.},
  \bibinfo{author}{Zhou, Y.}, \bibinfo{author}{Pasupathy, S.},
  \bibinfo{author}{Talwadker, R.}, \bibinfo{year}{2015}.
\newblock \bibinfo{title}{Hey, you have given me too many knobs!: Understanding
  and dealing with over-designed configuration in system software}, in:
  \bibinfo{booktitle}{Proceedings of the 2015 10th Joint Meeting on Foundations
  of Software Engineering}, \bibinfo{organization}{ACM}. pp.
  \bibinfo{pages}{307--319}.
\bibitem[{Xu et~al.(2003)Xu, Yu, Tresp, Xu and Wang}]{xu2003representative}
\bibinfo{author}{Xu, Z.}, \bibinfo{author}{Yu, K.}, \bibinfo{author}{Tresp,
  V.}, \bibinfo{author}{Xu, X.}, \bibinfo{author}{Wang, J.},
  \bibinfo{year}{2003}.
\newblock \bibinfo{title}{Representative sampling for text classification using
  support vector machines}, in: \bibinfo{booktitle}{European conference on
  information retrieval}, \bibinfo{organization}{Springer}. pp.
  \bibinfo{pages}{393--407}.
\bibitem[{Yarowsky(1995)}]{yarowsky1995unsupervised}
\bibinfo{author}{Yarowsky, D.}, \bibinfo{year}{1995}.
\newblock \bibinfo{title}{Unsupervised word sense disambiguation rivaling
  supervised methods}, in: \bibinfo{booktitle}{33rd annual meeting of the
  association for computational linguistics}, pp. \bibinfo{pages}{189--196}.
\bibitem[{Yu et~al.(2018)Yu, Bei and Qian}]{yu2018datasize}
\bibinfo{author}{Yu, Z.}, \bibinfo{author}{Bei, Z.}, \bibinfo{author}{Qian,
  X.}, \bibinfo{year}{2018}.
\newblock \bibinfo{title}{Datasize-aware high dimensional configurations
  auto-tuning of in-memory cluster computing}, in:
  \bibinfo{booktitle}{Proceedings of the Twenty-Third International Conference
  on Architectural Support for Programming Languages and Operating Systems},
  pp. \bibinfo{pages}{564--577}.
\bibitem[{Zhang et~al.(2018)Zhang, Van~Aken, Wang, Dai, Jiang, Lao, Sheng,
  Pavlo and Gordon}]{zhang2018demonstration}
\bibinfo{author}{Zhang, B.}, \bibinfo{author}{Van~Aken, D.},
  \bibinfo{author}{Wang, J.}, \bibinfo{author}{Dai, T.},
  \bibinfo{author}{Jiang, S.}, \bibinfo{author}{Lao, J.},
  \bibinfo{author}{Sheng, S.}, \bibinfo{author}{Pavlo, A.},
  \bibinfo{author}{Gordon, G.J.}, \bibinfo{year}{2018}.
\newblock \bibinfo{title}{A demonstration of the ottertune automatic database
  management system tuning service}.
\newblock \bibinfo{journal}{Proceedings of the VLDB Endowment}
  \bibinfo{volume}{11}, \bibinfo{pages}{1910--1913}.
\bibitem[{Zhang et~al.(2015)Zhang, Guo, Blais and
  Czarnecki}]{zhang2015performance}
\bibinfo{author}{Zhang, Y.}, \bibinfo{author}{Guo, J.}, \bibinfo{author}{Blais,
  E.}, \bibinfo{author}{Czarnecki, K.}, \bibinfo{year}{2015}.
\newblock \bibinfo{title}{Performance prediction of configurable software
  systems by fourier learning (t)}, in: \bibinfo{booktitle}{2015 30th IEEE/ACM
  International Conference on Automated Software Engineering (ASE)},
  \bibinfo{organization}{IEEE}. pp. \bibinfo{pages}{365--373}.
\bibitem[{Zhang et~al.(2016)Zhang, Guo, Blais, Czarnecki and
  Yu}]{zhang2016mathematical}
\bibinfo{author}{Zhang, Y.}, \bibinfo{author}{Guo, J.}, \bibinfo{author}{Blais,
  E.}, \bibinfo{author}{Czarnecki, K.}, \bibinfo{author}{Yu, H.},
  \bibinfo{year}{2016}.
\newblock \bibinfo{title}{A mathematical model of performance-relevant feature
  interactions}, in: \bibinfo{booktitle}{Proceedings of the 20th International
  Systems and Software Product Line Conference}, pp. \bibinfo{pages}{25--34}.
\bibitem[{Zheng et~al.(2014)Zheng, Ding and Hu}]{zheng2014self}
\bibinfo{author}{Zheng, C.}, \bibinfo{author}{Ding, Z.}, \bibinfo{author}{Hu,
  J.}, \bibinfo{year}{2014}.
\newblock \bibinfo{title}{Self-tuning performance of database systems with
  neural network}, in: \bibinfo{booktitle}{International Conference on
  Intelligent Computing}, \bibinfo{organization}{Springer}. pp.
  \bibinfo{pages}{1--12}.
\bibitem[{Zhou and Li(2005)}]{zhou2005tri}
\bibinfo{author}{Zhou, Z.H.}, \bibinfo{author}{Li, M.}, \bibinfo{year}{2005}.
\newblock \bibinfo{title}{Tri-training: Exploiting unlabeled data using three
  classifiers}.
\newblock \bibinfo{journal}{IEEE Transactions on knowledge and Data
  Engineering} \bibinfo{volume}{17}, \bibinfo{pages}{1529--1541}.
\bibitem[{Zhou and Li(2007)}]{zhou2007semisupervised}
\bibinfo{author}{Zhou, Z.H.}, \bibinfo{author}{Li, M.}, \bibinfo{year}{2007}.
\newblock \bibinfo{title}{Semisupervised regression with cotraining-style
  algorithms}.
\newblock \bibinfo{journal}{IEEE Transactions on Knowledge and Data
  Engineering} \bibinfo{volume}{19}, \bibinfo{pages}{1479--1493}.
\bibitem[{Zhou and Li(2010)}]{zhou2010semi}
\bibinfo{author}{Zhou, Z.H.}, \bibinfo{author}{Li, M.}, \bibinfo{year}{2010}.
\newblock \bibinfo{title}{Semi-supervised learning by disagreement}.
\newblock \bibinfo{journal}{Knowledge and Information Systems}
  \bibinfo{volume}{24}, \bibinfo{pages}{415--439}.
\bibitem[{Zhou et~al.(2007)Zhou, Zhan and Yang}]{zhou2007semi}
\bibinfo{author}{Zhou, Z.H.}, \bibinfo{author}{Zhan, D.C.},
  \bibinfo{author}{Yang, Q.}, \bibinfo{year}{2007}.
\newblock \bibinfo{title}{Semi-supervised learning with very few labeled
  training examples}, in: \bibinfo{booktitle}{AAAI}.
\bibitem[{Zhu et~al.(2003)Zhu, Ghahramani and Lafferty}]{zhu2003semi}
\bibinfo{author}{Zhu, X.}, \bibinfo{author}{Ghahramani, Z.},
  \bibinfo{author}{Lafferty, J.D.}, \bibinfo{year}{2003}.
\newblock \bibinfo{title}{Semi-supervised learning using gaussian fields and
  harmonic functions}, in: \bibinfo{booktitle}{Proceedings of the 20th
  International conference on Machine learning (ICML-03)}, pp.
  \bibinfo{pages}{912--919}.
\bibitem[{Zhu and Liu(2019)}]{zhu2019classytune}
\bibinfo{author}{Zhu, Y.}, \bibinfo{author}{Liu, J.}, \bibinfo{year}{2019}.
\newblock \bibinfo{title}{Classytune: A performance auto-tuner for systems in
  the cloud}.
\newblock \bibinfo{journal}{IEEE Transactions on Cloud Computing} .
\bibitem[{Zhu et~al.(2017)Zhu, Liu, Guo, Bao, Ma, Liu, Song and
  Yang}]{zhu2017bestconfig}
\bibinfo{author}{Zhu, Y.}, \bibinfo{author}{Liu, J.}, \bibinfo{author}{Guo,
  M.}, \bibinfo{author}{Bao, Y.}, \bibinfo{author}{Ma, W.},
  \bibinfo{author}{Liu, Z.}, \bibinfo{author}{Song, K.}, \bibinfo{author}{Yang,
  Y.}, \bibinfo{year}{2017}.
\newblock \bibinfo{title}{Bestconfig: tapping the performance potential of
  systems via automatic configuration tuning}, in:
  \bibinfo{booktitle}{Proceedings of the 2017 Symposium on Cloud Computing},
  pp. \bibinfo{pages}{338--350}.

\end{thebibliography}





\end{document}